\documentstyle[preprint,aps,eqsecnum,tighten,twoside]{revtex}

\def\be{\nopagebreak[3]\begin{equation}}
\def\ee{\end{equation}}
\def\ba{\nopagebreak[3]\begin{eqnarray}}
\def\ea{\end{eqnarray}}

\def\I{\it I}
\def\d{\partial}
\def\={\cong}
\def\L{{\cal L}}
\def\t{\tilde}
\def\I{\it I}
\def\d{\partial}
\def\={\cong}
\def\L{{\cal L}}
\def\S{{\cal S}}
\def\G{{\cal G}}
\def\B{{\cal B}}
\def\u{\underline}
\def\underbar{\u}

\begin{document}
\baselineskip=15pt
\title{Asymptotic structure of symmetry reduced general relativity}
\author{Abhay Ashtekar${}^{1}$, Ji\v r{\'\i} Bi\v c\'ak${}^{2}$,
and Bernd G. Schmidt${}^{3}$}
\address{${}^1$ Center for Gravitational Physics and Geometry\\
Department of Physics, Penn State, University Park, PA 16802, USA}
\address{${}^2$ Department of Theoretical Physics, Charles University\\
V Hole\v sovi\v ck\'ach 2, 180 00 Prague 8, Czech Republic}
\address{${}^3$ Max-Planck-Institut f{\"u}r Gravitationsphysik,\\
Schlaatzweg 1, 14473 Potsdam, Germany}
\pagenumbering{arabic}
\maketitle 

\pagenumbering{arabic}

\begin{abstract}

Gravitational waves with a space-translation Killing field are
considered. In this case, the 4-dimensional Einstein vacuum equations
are equivalent to the 3-dimensional Einstein equations with certain
matter sources. This interplay between 4- and 3- dimensional general
relativity can be exploited effectively to analyze issues pertaining
to 4 dimensions in terms of the 3-dimensional structures. An example
is provided by the asymptotic structure at null infinity: While these
space-times fail to be asymptotically flat in 4 dimensions, they can
admit a regular completion at null infinity in 3 dimensions. This
completion is used to analyze the asymptotic symmetries, introduce the
analog of the 4-dimensional Bondi energy-momentum and write down a
flux formula.

The analysis is also of interest from a purely 3-dimensional
perspective because it pertains to a diffeomorphism invariant
3-dimensional field theory with {\it local} degrees of freedom, i.e.,
to a midi-superspace. Furthermore, due to certain peculiarities of 3
dimensions, the description of null infinity does have a number of
features that are quite surprising because they do not arise in the
Bondi-Penrose description in 4 dimensions.
\end{abstract}

\section{Introduction}
\label{s1}

Einstein-Rosen waves are among the simplest non-stationary solutions
to the vacuum Einstein equations (see, e.g., \cite{1}). Not surprisingly,
therefore, they have been used in a number of different contexts:
investigation of energy loss due to gravity waves \cite{2}, asymptotic
structure of radiative space-times \cite{3}, quasi-local mass \cite{4}, the
issue of time in canonical gravity \cite{5}, and quantum gravity in a
simplified but field theoretically interesting context of
midi-superspaces \cite{5,6}. These solutions admit two Killing fields, both
hypersurface orthogonal, of which one is rotational, $\d/\d\phi$, and
the other translational, $\d/\d z$, along the axis of symmetry. (In
certain applications, the orbits of the Killing field
$\partial/\partial z$ are compactified, i.e., are taken to be circles.
Our analysis will allow this possibility.)  When the hypersurface
orthogonality condition is removed, we obtain the cylindrical
gravitational waves with {\it two} polarization modes. These have also
been used to explore a number of issues, ranging from the study of
Hamiltonian densities \cite{7} and numerical analysis of interacting pulses
\cite{8} to the issue of cosmic censorship \cite{9}.

The presence of a translational Killing field, however, makes the
analysis of the asymptotic structure of these space-times quite
difficult: they fail to be asymptotically flat either at spatial or
null infinity. Consequently, one can not use the standard techniques
to define asymptotic symmetries or construct the analogs of the ADM or
Bondi energy momenta. Therefore, until recently, conserved quantities
for these space-times --such as the C-energy \cite{2,7}-- were constructed
by exploiting the local field equations, without direct reference to
asymptotics. It is not apriori clear, therefore, that the quantities
have the physical interpretation that has been ascribed to them.

What is of physical interest are the values of conserved quantities
{\it per unit length} along the axis of symmetry, i.e. along the
integral curves of $\d/\d z$; because of the translational symmetry,
the total conserved quantities in such a space-time would be clearly
infinite.  A natural strategy then is to go to the manifold of orbits
of the $\d/\d z$-Killing field. Since this 3-dimensional space-time
does not have a translational symmetry, one would expect it to be
asymptotically flat in an appropriate sense. Hence, it should be
possible to analyze its asymptotic structure unambiguously.  In this
paper, we will adopt this approach to explore the symmetries and
physical fields at null infinity. A similar analysis of spatial
infinity was performed recently \cite{10} in the context of the phase space
formulation of general relativity. Somewhat surprisingly, it turned
out that the C-energy is {\it not} the generator of the time
translation which is unit at infinity; it does not therefore represent
the Hamiltonian, or the physical energy (per unit {\it z}-length) in
the space-time. The physical Hamiltonian turns out to be a {\it
non-polynomial} function of the C-energy. In the present paper, we
will see that the same is true of the analog of Bondi energy at null
infinity.

Thus, the purpose of this paper is to develop a framework to discuss
the asymptotic structure at null infinity for 3-dimensional
space-times.  The underlying theory is general relativity coupled to
matter fields satisfying appropriate fall-off conditions. The
conditions on matter are satisfied, in particular, by the fields that
arise from a symmetry reduction of a large class of 4-dimensional
vacuum space-times admitting a space translation. Therefore, we will,
in particular, provide a framework for analyzing the behavior of the
gravitational field near null infinity of such space-times.  We call
such space-times generalized cylindrical waves since they need
not admit an axial Killing field $\d/\d\phi$. Our analysis is also
useful in a completely different context; that of quantum
gravity. For, this class of space-times also provides interesting
midi-superspace for quantum gravity and our results set the stage for
its asymptotic quantization and the corresponding S-matrix theory.

The plan of the paper is as follows. In Sec.\ref{s2}, we will analyze
the asymptotic structure of the Einstein-Rosen waves from a
3-dimensional perspective. This analysis will motivate our general
definition of asymptotic flatness in Sec.\ref{s3} and also make the
main results plausible. In Sec.\ref{s3}, we introduce the notion of
asymptotic flatness at null infinity in 3 space-time dimensions and
analyze the structure of asymptotic fields. In Sec.\ref{s4}, we
discuss asymptotic symmetries and in Sec.\ref{s5}, conserved
quantities. While the general methods adopted are suggested by the
standard Bondi-Penrose treatment of null infinity in 4-dimensional
general relativity, there are a number of surprises as well. First, in
3 dimensions, the physical metric $g_{ab}$ is {\it flat} outside
sources.  Consequently, there are physically interesting solutions to
the constraints which lead to space-times which are flat near spatial
infinity $i^o$; the energy-momentum at $i^o$ is coded, not in local
fields such as the curvature, but in a globally defined deficit angle.
This simplifies the task of specifying boundary conditions as one
approaches $i^o$ along null infinity $\I$. On the other hand, there
are also a number of new complications.  In 4 dimensions, the
stationary and the radiative space-times satisfy the same boundary
conditions at null infinity. This is not the case in 3 dimensions.
Hence, while dealing with radiative solutions, we can not draw on our
intuition from the stationary case. Secondly, in 4 dimensions, up to a
super-translation freedom --which corresponds to terms $O(1/r)$--
there is a fixed Minkowskian metric at infinity. In 3 dimensions, this
is not the case; the Minkowski metric $\eta_{ab}$ to which a physical
metric approaches varies even in the leading order, depending on the
radiative content of the physical space-time.  Consequently, the
symmetry group is larger than what one might expect from one's
experience in 4 dimensions. Furthermore, while one can canonically
single out the translational subgroup of the BMS group in 4
dimensions, now the task becomes subtle; in many ways it is analogous
to the task of singling out a preferred Poincar\'e subgroup of the BMS
group. This in turn makes the task of defining the analog of Bondi
energy much more difficult.  These differences make the analysis
non-trivial and hence interesting.

Some detailed calculations are relegated to appendices. Using
Bondi-type coordinates, the asymptotic behavior of curvature tensors
of Einstein-Rosen waves is analyzed in the 3-dimensional framework in
Appendix A. Appendix B considers static cylindrical solutions whose
asymptotics, as mentioned above, is quite different from that of the
radiative space-times analyzed in the main body of the paper.

It should be emphasized that while part of the motivation for our
results comes from the symmetry reduction of 4-dimensional general
relativity, the main analysis itself refers to 3-dimensional gravity
coupled to {\it arbitrary} matter fields (satisfying suitable
fall-off conditions) which need not arise from a symmetry
reduction. Nonetheless, the framework has numerous applications to the
4-dimensional theory. For example, in the accompanying paper
\cite{16}, we will use the results of this paper to study the behavior
of Einstein-Rosen waves at null infinity of the {\it 4-dimensional}
space-times.

In this paper, the symbol $\I$ will generally stand for $\I^+$ or
$\I^-$. In the few cases where a specific choice has to be made, our
discussion will refer to $\I^+$.

\goodbreak  

\section{Einstein-Rosen waves: Asymptotics in 3 dimensions}
\label{s2}

This section is divided into three parts. In the first, we recall the
symmetry reduction procedure and apply it to obtain the 3-dimensional
equations governing Einstein-Rosen waves. (See, e.g., \cite{1} for a
similar reduction for stationary space-times.)  This procedure reduces
the task of finding a 4-dimensional Einstein-Rosen wave to that of
finding a solution to the wave equation on 3-dimensional {\it
Minkowski} space.  In the second part, we analyze the asymptotic
behavior (at null infinity) of these solutions to the wave equation.
In the third part, we combine the results of the first two to analyze
the asymptotic behavior of space-time metrics. We will find that there
is a large class of Einstein-Rosen waves which admit a smooth null
infinity, $\I$, as well as a smooth time-like infinity $i^\pm$. (As
one might expect, the space-like infinity, $i^o$, has a conical
defect.)  These waves provide an important class of examples of the
more general framework presented in Sec.\ref{s3}.

\goodbreak
\subsection{Symmetry reduction}
\label{s2.1}

Let us begin with a slightly more general context, that of vacuum
space-times which admit a space-like, hypersurface orthogonal Killing
vector $\partial/\partial z$. These space-times can be described
conveniently in coordinates adapted to the symmetry:
\be
ds^2 = V^2(x)dz^2 +\bar g_{ab}(x)\ dx^adx^b\ ,\ \ \ a,b,\dots =0,1,2
\label{(2.1)}
\ee
where $x \equiv x^a $ and $\bar g_{ab}$ is a 3-metric metric with
Lorentz signature. As in the more familiar case of static space-times
\cite{1} the field equations are
\be
\bar R_{ab}-V^{-1}\bar\nabla_a \bar\nabla_b V =0\ ,\ \  
\label{(2.2)}
\ee
\be
\bar g^{ab}  \bar\nabla_a \bar\nabla_b V =0 \ ,
\label{(2.3)}
\ee 
where $\bar\nabla$ and $\bar R_{ab}$ are the derivative operator and
the Ricci tensor of $\bar g_{ab}$. These equations can be simplified
if one uses a metric in the 3-space which is rescaled by the norm of
the Killing vector and writes the norm of the Killing vector as an
exponential \cite{12,1}. Then (2.1)--(2.3) become 
\be 
ds^2 = e^{2\psi(x)}dz^2 +e^{-2\psi(x)} g_{ab}(x)\ dx^adx^b\ ,
\label{(2.4)}      
\ee
\be
 R_{ab}-2 \nabla_a \psi \nabla_b \psi =0\ ,
\label{(2.5)}
\ee
\be
{} \quad g^{ab}\nabla_a\nabla_b \psi=0 \ ,
\label{(2.6)}
\ee 
where $\nabla$ denotes the derivative with respect to the metric
$g_{ab}$.

These equations can be re-interpreted purely in a 3-dimensional
context. To see this, consider Einstein's equations in 3 dimensions
with a scalar field $\Phi$ as source:
\be
R_{ab}-{1\over 2}R\, g_{ab}= 8\pi G T_{ab} =8\pi G (\nabla_a \Phi
\nabla_b\Phi - \textstyle{1\over 2} (\nabla_c\Phi
\nabla^c\Phi)\, g_{ab})\ ,
\label{(2.7)}
\ee 
\be
g^{ab}\nabla_a \Phi \nabla_b\Phi = 0\ . \label{(2.8)}
\ee
Since the trace of equation (2.7) gives $R = 8\pi G\nabla^c\Phi
\nabla_c\Phi$, (2.7) is equivalent to 
\be 
R_{ab}=  8\pi G \, \nabla_a\Phi \nabla_b\Phi \ .
\label{(2.9)}
\ee 
Now, with $\Phi = \psi/\sqrt{4\pi G} $ we obtain (2.5) and
(2.6). Thus, the 4-dimensional vacuum gravity is equivalent to the
3-dimensional gravity coupled to a scalar field. Recall that in 3
dimensions, there is no gravitational radiation. Hence, the local
degrees of freedom are all contained in the scalar field.  One
therefore expects that the Cauchy data for the scalar field will
suffice to determine the solution. For data which fall off
appropriately, we thus expect the 3-dimensional Lorentzian geometry to
be asymptotically flat in the sense of Penrose \cite{13}, i.e. to
admit a 2-dimensional boundary representing null infinity.

Let us now turn to the Einstein-Rosen waves by assuming that there is
further space-like, hypersurface orthogonal Killing vector $\d/\d\phi$
which commutes with $\partial/ \partial z$. Then, as is well known,
the equations simplify drastically. Hence, a complete global analysis
can be carried out easily.  Recall first that the metric of a vacuum
space-time with two commuting, hypersurface orthogonal space-like
Killing vectors can always be written locally as \cite{14}
\be 
ds^2=e^{2\psi}dz^2+ e^{2(\gamma
-\psi)}(-dt^2+d\rho^2)+\rho^2e^{-2\psi}d\phi^2\ ,
\label{(2.10)} 
\ee 
where $\rho$ and $t$ (the ``Weyl canonical coordinates'') are defined
invariantly and $\psi=\psi(t,\rho)$, $\gamma=\gamma (t,\rho)$.  (Here,
some of the field equations have been used.) Hence the 3-metric $g$ is
given by
\be 
d\sigma^2=g_{ab}dx^adx^b=e^{2\gamma}(-dt^2+d\rho^2)
+\rho^2d\phi^2\ .
\label{(2.11)} 
\ee 
Let us now assume that $\d/\d\phi$ is a rotational field in the
3-space which keeps a time-like axis fixed.  Then the coordinates used
in (2.10) are unique up to a translation $t\rightarrow t+a$. (Note,
incidentally, that ``trapped circles'' are excluded by the field
equations \cite{9}.)

The field equations (\ref{(2.5)}) and (\ref{(2.6)}) now become
\ba
R_{tt}&=&\gamma''-\ddot\gamma +\rho^{-1}\gamma'=2\dot\psi^2\ ,
\\
R_{\rho\rho}&=& -\gamma''+\ddot\gamma +\rho^{-1}\gamma'=2\psi'^2\ ,
\\
R_{t\rho}&=& \rho^{-1}\dot\gamma =2\dot\psi\psi'\ ,
\ea
and
\be
-\ddot\psi+\psi''+ \rho^{-1}\psi'=0 \ ,
\label{(2.15)}
\ee
where the dot and the prime denote derivatives with respect to $t$ and
$\rho$ respectively.  The last equation is the wave equation for the
non-flat 3-metric (2.11) {\it as well as for the flat metric obtained
by setting} $\gamma=0$.  This is a key simplification for it implies
that the equation satisfied by the matter source $\psi$ decouples from
the equations (2.12)-(2.14) satisfied by the metric. These equations
reduce simply to: \be
\gamma ' =\rho\,(\dot\psi^2+\psi'^2)\ ,
\label{(2.16)}
\ee
\be
\dot\gamma = 2\rho\dot\psi\psi '\ .
\label{(2.17)}
\ee 
Thus, we can first solve for the axi-symmetric wave equation (2.15)
for $\psi$ on Minkowski space and then solve (2.16) and (2.17) for
$\gamma$ --the only unknown metric coefficient-- by quadratures.
(Note that (2.16) and (2.17)  are compatible because their
integrability condition is precisely (2.15).)
\goodbreak
\subsection{Asymptotic behavior of scalar waves}
\label{s2.2}

In this subsection we will focus on the axi-symmetric wave
equation in 3-dimen\-sion\-al Minkowski space and analyze the
asymptotic behavior of its solutions $\psi$.

We begin with an observation.  The ``method of descent'' from the
Kirchoff formula in 4 dimensions gives the following representation
of the solution of the wave equation in 3 dimensions, in terms of
Cauchy data $\Psi_0=\psi(t=0,x,y), \Psi_1=\psi_{,t}(t=0,x,y)$: 
\ba
\psi(t,x,y)&=& {1\over 2 \pi}\ {\partial\over \partial t}
\int\!\!\!\int_{S(t)}
{\Psi_0(x',y')dx'dy'
\over [t^2-(x-x')^2-(y-y')^2]^{1/2}}
\nonumber\\
& & +{1\over 2 \pi}
\int\!\!\!\int_{S(t)}
{\Psi_1(x',y')dx'dy'
\over [t^2-(x-x')^2-(y-y')^2]^{1/2}}\ ,
\label{(2.18)}
\ea
where $S$ is the disk 
\be
(x-x')^2+(y-y')^2\le t^2
\nonumber\ee
in the initial Cauchy surface (see, e.g., \cite{29}).  We will assume
that the Cauchy data are axially symmetric and of compact support.

Let us first investigate the behavior of the solution at future null
infinity $\I$. Let $\rho,\phi$ be polar coordinates in the plane and
introduce the retarded time coordinate
\be
u=t-\rho
\nonumber\ee 
to explore the fall-off along the constant $u$ null hypersurfaces.
Because of axi-symmetry, we may put $y=0$ without loss of
generality.  The integration region becomes
\be
(\rho -x')^2+y'^2\le (u+\rho)^2 . 
\nonumber\ee
Let us rewrite the integrands of (\ref{(2.18)}) as
\be
{\Psi(x',y')dx'dy'
\over [2\rho(u+x')+u^2-x'^2-y'^2]^{1/2}}
={1\over{(2\rho)}^{1/2}}{\Psi(x',y')dx'dy'\over(u+x')^{1/2}}
\left(1+{u^2-x'^2-y'^2\over2(u+x')}\ {1\over\rho}\right)^{-1/2}\ .
\label{(2.22)}
\ee
For large $\rho$, (\ref{(2.22)}) admits a power series expansion in
$\rho^{-1}$ which converges absolutely and uniformly. Hence we can
exchange the integration in (\ref{(2.18)}) with the summation and we
can also perform the differentiation $\partial/ \partial u $ term by
term.  Therefore on each null hypersurface $u=const$ one can obtain an
expansion of the form
\be
\psi(u,\rho)={1\over\sqrt\rho}\left(f_0(u)+\sum_{k=1}^\infty{f_k(u)
\over\rho^k}\right)\ .
\label{(2.23)} 
\ee

The coefficients in this expansion are determined by integrals over
the Cauchy data. These functions are particularly interesting for $u$
so large that the support of the data is completely in the interior of
the past cone. One finds
\be
f_0(u)={1\over 2\sqrt2\pi}\int_0^\infty\!\int_0^{2\pi}\rho'd\rho'd\phi'
\left[-{1\over2}{\Psi_0\over(u+\rho'\cos\phi')^{3\over2}}
      +{\Psi_1\over(u+\rho'\cos\phi')^{1\over2}}
\right]\ .
\label{(2.24)}
\ee
Note that the coefficient is analytic in $u^{-1/2}$, and at $u\gg
\rho_0$, $\rho_0$ being the radius of the disk in which the data are
non-zero, we obtain
\be
f_0(u)={k_0\over u^{3/2}}+{k_1\over u^{1/2}}+\dots\ ,
\label{(2.25)}
\ee 
where $k_0,k_1$ are constants which are determined by the data.  If
the solution happens to be time-symmetric, so that $\Psi_1$ vanishes,
we find $f_0\sim u^{-3/2}$ for large $u$. This concludes our discussion
of the asymptotic behavior along $u = const$ surfaces.

Finally, we wish to point out that the main results obtained in this
section continue to hold also for general data of compact support
which are not necessarily axi-symmetric. In particular, one obtains an
expansion like (\ref{(2.23)}) where the coefficients now depend on
both $u$ and $\phi$, and asymptotic forms like (\ref{(2.25)}).  The
assumption of compact support can also be weakened to allow data which
decay near spatial infinity sufficiently rapidly so that we still
obtain solutions smooth at null infinity. (This is in particular the
case for the Weber-Wheeler pulse considered in the accompanying paper
\cite{16}.)
\goodbreak
\subsection{Asymptotic behavior of the metric}
\label{s2.3}

We now combine the results of the previous two subsections. Recall
from Eq. (\ref{(2.11)}) that the 3-dimensional metric $g_{ab}$ has a
single unknown coefficient, $\gamma(t, \rho)$, which is determined by
the solution $\psi(t, \rho)$ to the wave equation in Minkowski space
(obtained simply by setting $\gamma= 0$).  The asymptotic behavior of
$\psi(t,\rho)$ therefore determines that of the metric $g_{ab}$.

Let us begin by expressing $g_{ab}$ in the Bondi-type coordinates
$(u=t-\rho ,\rho,\phi)$. Then, Eq. (\ref{(2.11)}) yields
\be
d\sigma^2=e^{2\gamma}(-du^2-2du d\rho) + \rho^2d\phi^2\ ;
\label{(2.26)} 
\ee
the Einstein equations take the form
\be
\gamma_{,u}=2\rho\psi_{,u}(\psi_{,\rho}-\psi_{,u})\ ,
\label{(2.27)}
\ee
\be
\gamma_{,\rho}=\rho\psi^2_{,\rho}\ ;
\label{(2.28)}
\ee
and the wave equation on $\psi$ becomes
\be
-2\psi_{,u\rho}+\psi_{,\rho\rho} +\rho^{-1}(\psi_{,\rho}-\psi_{,u})=0\ .
\label{(2.29)}
\ee 
The asymptotic form of $\psi(t,\rho)$ is given by the expansion
(\ref{(2.23)}).  Since we can differentiate (\ref{(2.23)}) term by
term, the field equations (\ref{(2.27)}) and (\ref{(2.28)}) imply
\be
\gamma_{,u}= -2[\dot f_0(u)]^2+\sum_{k=1}^\infty{g_k(u)\over\rho^k}\ ,
\label{(2.30)}
\ee
\be
\gamma_{,\rho}= {1\over 4}[\dot f_0(u)]^2\ {1\over {\rho^2}}
+\sum_{k=1}^\infty{h_k(u)\over\rho^{k+2}}\ ,
\label{(2.31)}
\ee
where the functions $g_k, h_k$ are products of the functions $f_0,
f_k, \dot f_0, \dot f_k$. Since for data of compact support $f_0, f_k$
vanish for all $u<u_0$, we can integrate (\ref{(2.30)}) as follows:
\be
\gamma = \gamma_0 + \int_{-\infty}^u\left( -2(\dot f_0(u))^2+
\sum_{k=1}^\infty{g_k(u)\over\rho^k}\right)du\ .
\label{(2.32)}
\ee 
Thus, $\gamma$ also admits an expansion in $\rho^{-1}$ where
the coefficients depend smoothly on $u$.

It is now straightforward to show that the space-time admits a smooth
future null infinity, $\I$.  Setting $\tilde\rho=\rho^{-1}, \tilde
u=u, \tilde\phi =\phi$ and  rescaling $g_{ab}$ by a conformal factor
$\Omega=\tilde\rho $, we obtain
\be
d\tilde \sigma^2=\Omega^2 d\sigma^2=e^{2\tilde\gamma}
(-\tilde\rho^2d\tilde u^2+2d\tilde
ud\tilde\rho)+d\tilde\phi^2\ ,  
\label{(2.33)}
\ee 
where $\tilde\gamma(\tilde u, \tilde\rho)=\gamma(u, \tilde\rho^{-1})$.
Because of (\ref{(2.32)}), $\tilde\gamma$ has a smooth extension through
$\tilde\rho=0$. Therefore, $\tilde{g}_{ab}$ is smooth across the
surface $\tilde\rho= 0$. This surface is the future null infinity,
$\I$.

Using the expansion (\ref{(2.23)}) of $\psi$ near null infinity,
various curvature tensors can be expanded in powers of
$\rho^{-1}$. More precisely, a suitable null triad can be chosen which
is parallel propagated along $u=const$, $\phi=const$ curves. The
resulting triad components of the Riemann tensor and the Bach tensor
are given in Appendix A. The (conformally invariant) Bach tensor is
finite {\it but non-vanishing} at null infinity. This is to be
contrasted with the Bondi-Penrose description of null infinity in
asymptotically flat, 4-dimensional space-times, where the (conformally
invariant) Weyl tensor vanishes. In this sense, while in the standard
4-dimensional treatments the metric is conformally flat {\it at} null
infinity, in a 3-dimensional treatment, it will not be so in
general. This is one of the new complications that one encounters.

To understand the meaning of the constant $\gamma_0$ let us consider
the solution on the Cauchy surface $t=0$. Eq. (\ref{(2.16)}) implies
that we can determine $\gamma$ by a $\rho$-integration from the
center.  If we insist on regularity at $\rho =0$ we have
\be
\gamma(t=0,\rho)=\int_0^\rho \rho\,(\dot\psi^2+\psi'^2)\ d\rho\ .
\label{(2.34)} 
\ee 
Hence, for data of compact support, $\gamma_0$ is a positive constant
whose value is determined by the initial data for $\psi$:
\be
\gamma_0=\gamma(t=0,\rho=\infty)=\int_0^\infty \rho\,(\dot\psi^2+
\psi'^2)\ d\rho \ .
\label{(2.35)} 
\ee
This way the constant $\gamma_0$ in (\ref{(2.32)}) is uniquely
determined for solutions which are regular at $\rho=0$. Its value is
given by the total energy of the scalar field $\psi$ computed using
the Minkowski metric (obtained from $g_{ab}$ by setting $\gamma =0$).

On a constant $t$ surface, for a point outside the support of the
data, we have $\gamma =\gamma_0$, a constant. Hence, outside the
support of the data, the 3-metric on the Cauchy surface is flat. For
any non-trivial data, however, $\gamma_0$ is strictly positive, whence
the metric has a ``conical singularity'' at spatial infinity: the
metric there is given by
\be
d\sigma^2=e^{2\gamma_0}(-dt^2+d\rho^2)+\rho^2d\phi^2\ .
\label{(2.36)} 
\ee
Notice that a conical singularity can also be seen near null infinity
in this physical metric because the change of the proper circumference
of a circle with proper radial distance is different from the case of
asymptotically Minkowskian space.

Finally, using (\ref{(2.32)}), we find that, as one approaches $\I$
(i.e. $\rho\to\infty$), we have:
\be
\gamma(u,\infty)= \gamma_0 -2\int_{-\infty}^u \dot f_0(u)^2du .
\label{(2.37)}
\ee 
Now, a detailed examination \cite{16} of the behavior of the scalar
field $\psi$ near time-like infinity $i^+$ reveals that the space-time
is smooth at ${i}^+$ and that $\gamma$ vanishes there. Hence, we
obtain the simple result
\be
\gamma_0 =2\int_{-\infty}^{+\infty} \dot f_0(u)^2du .
\label{(2.38)}
\ee
Thus, there, is a precise sense in which the conical singularity,
present at space-like infinity, is ``radiated out'' and a smooth (in
fact analytic) time-like infinity ``remains". We will see that, modulo
some important subtleties, Eq. (\ref{(2.37)}) plays the role of the
Bondi mass-loss formula \cite{15}.
\goodbreak

\section{ Null infinity in 3 dimensions: general framework}
\label{s3}

In this section, we will develop a general framework to analyze the
asymptotic structure of the gravitational and matter fields at null
infinity in 3 dimensions along the lines introduced by Penrose in 4
dimensions. As a special case, when the matter field is chosen to be
the massless Klein-Gordon field, we will recover a 3-dimensional
description of null infinity of generalized cylindrical waves. It
turns out that the choice of the fall-off conditions on matter fields
is rather subtle in 3 dimensions. Fortunately, the analysis of the
Einstein-Rosen waves presented in Sec.\ref{s2} provides guidelines
that restrict the available choices quite effectively.

In Sec.\ref{s3.1}, we specify the boundary conditions and discuss
some of their immediate consequences.  In \ref{s3.2}, we extract the
important asymptotic fields and discuss the equations they satisfy at
null infinity. Sec.\ref{s3.3} contains an example which, so to say,
lies at the opposite extreme from the Einstein-Rosen waves: the
simplest solution corresponding to a static point particle in
3 dimensions. This example is simple enough to bring out certain
subtleties which in turn play an important role in the subsequent
sections.
\goodbreak
\subsection{Boundary conditions}
\label{s3.1}

A 3-dimensional space-time $(M, g_{ab})$ will be said to be {\it
asymptotically flat at null infinity} if there exists a manifold
$\t{M}$ with boundary $\I$ which is topologically $S^1 \times R$,
equipped with a smooth metric $\t{g}_{ab}$ such that
\begin{itemize}
\item[i)]{there is a diffeomorphism between $\t{M} - \I$ and $M$ 
(with which we will identify the interior of $\t{M}$ and $M$);}
\item[ii)]{there exists a smooth function $\Omega$ on $\t{M}$  such 
that, at $\I$, we have  $\Omega = 0$, $\nabla_a \Omega \not= 0$, and 
on $M$, we have $\t{g}_{ab} =\Omega^2 g_{ab}$;} 
\item[iii)]{If $T_{ab}$ denotes the stress-energy of matter fields on
the physical space-time $(M, g_{ab})$, then $\Omega T_{ab}$ admits a
smooth limit to $\I$ which is {\it trace-free}, and the limit to $\I$
of $\Omega^{-1}T_{ab}\t{n}^a \t{V}^b$ vanishes, where $\t{V}^a$ is any
smooth vector field on $\t{M}$ which is tangential to $\I$ and
$\t{n}^a = \t{g}^{ab}\t\nabla_b\Omega$;}
\item[iv)]{if $\Omega$ is so chosen that $\t\nabla^a \t\nabla_a 
\Omega = 0$ on $\I$, then the vector field $\t{n}^a$ is 
complete on $\I$.}
\end{itemize}

Conditions i), ii) and iv) are the familiar ones from 4 dimensions and
have the following implications. First, since $\Omega$ vanishes at
$\I$, points of $\I$ can be thought of as lying at infinity with
respect to the physical metric. Second, since the gradient of $\Omega$
is non-zero at $\I$, $\Omega$ ``falls off as $1/\rho$''. Finally, we
know that $\I$ has the topology $S^1\times R$ and condition iv)
ensures that it is as ``complete in the $R$-direction'' as it is in
Minkowski space. 

The subtle part is the fall-off conditions on stress-energy; these are
{\it substantially weaker} than those in the standard 4-dimensional
treatment. For instance, in 4 dimensions, if we use Maxwell fields as
sources, then because of conformal invariance, if $F_{ab}$ solves
Maxwell's equations on the physical space-time $(M, g_{ab})$, then
$\t{F}_{ab} := F_{ab}$ satisfies them on the completed space-time
$(\t{M}, \t{g}_{ab})$. Hence $\t{F}_{ab}$ admits a smooth limit to
$\I$. This immediately implies that $\Omega^{-2} T_{ab}$ also admits a
smooth limit, where $T_{ab}$ is the stress-energy tensor of $F_{ab}$
in the physical space-time. In the case of a scalar field source, the
fall-off is effectively the same although the argument is more subtle
(see page 41 in \cite{17}). In 3 dimensions, on the other hand, we are
asking only that $\Omega T_{ab}$ admits a limit (although, as noted
above, the asymptotic fall-off of $\Omega$ is the same in 3 and 4
dimensions). This is because a stronger condition will have ruled out
the cylindrical waves discussed in Sec.\ref{s2}. To see this, consider
smooth scalar fields $\psi$ with initial data of compact
support. Then, if we set $\t\psi = \Omega^{-1/2}\psi$, we have the
identity:
$$ \t{g}^{ab} \t\nabla_a \t\nabla_b \t\psi - {\textstyle{1\over 8}}\,
\t{R} \t\psi = \Omega^{-{\textstyle{5\over 2}}}(g^{ab}\nabla_a\nabla_b 
\psi -{\textstyle{1\over 8}}\, R \psi)\ ,$$
where $R$ and $\t{R}$ are the scalar curvatures of $g_{ab}$ and
$\t{g}_{ab}$ respectively. Hence $\t\psi$ is well-behaved on $\I$
which implies that 
\ba\Omega T_{ab} &\equiv& 2 \Omega^2 (\t\nabla_a\t\psi) 
(\t\nabla_b \t\psi) +2 \Omega\t\psi \t{n}_{(a}\t\nabla_{b)}\t\psi 
+ \textstyle{1\over 2} \t{\psi}^2\t{n}_a \t{n}_b \nonumber\\
&-& \textstyle{1\over 2}\t{g}_{ab} [ \Omega^2
\t{\nabla}^m {\t\nabla}_m \t\psi + \Omega\t\psi \t{n}^m\t\nabla_m 
\t\psi + \t{n}^m \t{n}_m \t{\psi}^2]\label{se} \ea 
admits a well-defined, non-zero limit at $\I$ satisfying the
conditions of our definition. Hence, stronger fall-off requirements on
$T_{ab}$ would have made the framework uninteresting. We will see that
this weak fall-off is responsible for a number of surprises in the
3-dimensional theory.  Could we have imposed even weaker fall-off
conditions? The requirement of smoothness on $\t{g}_{ab}$, $\Omega$
and $\Omega T_{ab}$ can be substantially weakened: All our analysis
will go through if $\t{g}_{ab}$ and $\Omega$ are only $C^3$, and
$\Omega T_{ab}$ only $C^1$ at $\I$. On the other hand, we will see
that the condition on the trace of $\Omega T_{ab}$ is necessary to
endow $\I$ with interesting structure. We will see that the vanishing
of the limit of $\Omega^{-1} T_{ab} \t{n}^a \t{V}^b$ is necessary to
ensure that the energy and super-momentum fluxes of matter across
(finite patches of) $\I$ are finite.

Let us now examine the structure available at the boundary $\I$. 

As in 4 dimensions, it is convenient to work entirely with the
tilde fields which are smooth at $\I$. Let us set 
$$\t{L}_{ab} = \Omega (R_{ab} - \textstyle{1\over 4}R\, 
g_{ab})=: \Omega S_{ab} $$
and lower and raise its indices with $\t{g}_{ab}$ and its inverse.
$\t{L}_{ab}$ carries the same information as the stress-energy tensor
$T_{ab}$ of matter and our conditions on $T_{ab}$ ensure that
$\t{L}_{ab}$ is smooth at $\I$.  Set
$$\bar{f} =  \Omega^{-1} \t{n}^a \t{n}_a\, .  $$
Then, using the expression $R_{abcd} = 2 (S_{a[c}g_{d]b} - S_{b[c}
g_{d]a})$ of the Riemann tensor in 3 dimensions, the formula
expressing the relation between curvature tensors of $g_{ab}$ and
$\t{g}_{ab}$ reduces to:
\be\Omega \t{S}_{ab} + \t\nabla_a \t{n}_b - \textstyle{1\over 2} 
\bar{f} \t{g}_{ab} = \t{L}_{ab}\ , \label{(3.1)}\ee
where $\t{S}_{ab} = (\t{R}_{ab} - \textstyle{1\over 4}\t{R}\t{g}_{ab}
)$.  This is the basic field equation in the tilde variables. Since
all other fields which feature in it are known to be smooth at $\I$,
it follows that $\bar f$ is also smooth. This implies in particular
that $\t{n}^a$ is null. Since $\t{n}_a =
\t{\nabla}_a\Omega$ is the normal field to $\I$, we conclude that  
${\I}$ {\it is a null surface}.

Next, we note that there is a considerable freedom in the choice of
the conformal factor $\Omega$. Indeed, if $(\t{M}, \t{g}_{ab} =
\Omega^2 g_{ab})$ is an allowable completion, so is $(\t{M},
\Omega'^2 g_{ab})$ where $\Omega' = \omega\Omega$ for any smooth, 
nowhere vanishing function $\omega$ on $\t{M}$.  Now, under the
conformal transformation $\Omega\mapsto \Omega' = \omega\Omega$, we
have:
$$\t\nabla'_a \t{n}'^a \= \omega^{-1}\t\nabla_a \t{n}^a + 3 \omega^{-2}
\L_{\t{n}} \omega \ ,$$
where, from now on, $\=$ will stand for {\it ``equals at the points of
$\I$ to''}.  Hence, by using an appropriate $\omega$, we can always
make $\t{n}'^a$ divergence-free. Such a choice will be referred to as
a {\it divergence-free conformal frame}. This frame is, however, not
unique.  The restricted gauge freedom is given by:
\be\Omega \mapsto \omega\Omega,  \quad{\rm where}\quad \L_{\t{n}} 
\omega\= 0\ . \label{(3.2)}\ee
Now, condition iv) in our definition requires that, in any
divergence-free conformal frame, the vector field $\t{n}^a$ be
complete on $\I$. Suppose it is so in one divergence-free conformal
frame $\Omega$. Let $\Omega'$ correspond to another divergence-free
frame. Then, $\Omega' = \omega\Omega$, with $\omega$ smooth, nowhere
vanishing and satisfying $\L_{\t{n}} \omega \= 0$. The last equation
implies that $\t{n}'^a$ is complete on $\I$ if and only if $\t{n}^a$
is complete there. Hence, we need to verify iv) in just one
divergence-free conformal frame. {\it In what follows, we will work
only in divergence-free conformal frames.}

Next, taking the trace of (3.1) and using the fact that $\t{L}$
vanishes on $\I$ we conclude that, in any divergence-free frame,
$\bar{f}$ vanishes on $\I$, whence
$$ \t{f} := \Omega^{-1} \bar{f} $$ 
admits a smooth limit there.  The field $\t{f}$ will play an important
role. Finally, it is easy to check that in any divergence-free
conformal frame, we have:
\be\t{n}^b \t\nabla_b \t{n}_a \= 0,\quad {\rm and} \quad 
\t{L}_{ab} \t{n}^b \= 0\ . \label{(3.3)}\ee
Thus, in particular, as in 4 dimensions, $\I$ is ruled by null
geodesics. The space ${\B}$ of orbits of $\t{n}^a$ --the ``base space''
of $\I$-- is diffeomorphic to $S^1$.  The second equation and the
trace-free character of $\t{L}_{ab}$ imply that, {\it on} $\I$,
$\t{L}_{ab}$ has the form
\be\t{L}_{ab} \= \t{L}_{(a}\t{n}_{b)}\, , \quad{\rm with}\quad
\t{L}_a\t{n}^a \= 0\, ,  \label{(3.4)}\ee
for some smooth co-vector field $\t{L}_a$. Hence, the pull-back to
$\I$ of $\t{L}_{ab}$ vanishes which in turn implies, via (3.1), the
pull-back to $\I$ of $\t{\nabla}_a\t{n}_b$ also vanishes. Hence, if we
denote by $\t{q}_{ab}$ the pull-back of $\t{g}_{ab}$, we have:
\be\L_{\t{n}}  \t{q}_{ab} \=  0\ .\label{(3.5)} \ee
Since $\I$ is null, it follows that 
\be \t{q}_{ab}\t{n}^b \= 0\, .\label{ (3.6)} \ee
Thus, $\t{q}_{ab}$ is the pull-back to $\I$ of a positive definite
metric on the manifold of orbits $\B$ of the vector field
$\t{n}^a$. By construction, $\B$ is a 1-dimensional manifold with
topology of $S^1$.  Hence, there exists a 1-form $\t{m}_a$ on $\I$
such that
\be \t{q}_{ab} = \t{m}_a \t{m}_b\, . \label{(3.7)}\ee
(In cylindrical waves, $\t{m}_a$ is the pull-back to $\I$ of $\t
\nabla_a \phi$ and $\t{n}^a$ equals $\exp (-2\t{\gamma})\, (\partial/
\partial u)$ on $\I$.) Under a conformal rescaling $\Omega \mapsto 
\omega\Omega$ (from one divergence-free frame to another), we have:
\be \t{m}_a \mapsto \omega \t{m}_a  \quad \t{n}^a \mapsto 
\omega^{-1} \t{n}^a\, .\label{(3.8)}  \ee
The pairs $(\t{m}_{a}, \t{n}^a)$ (or, equivalently, $(\t{q}_{ab},
\t{n}^a))$ are the kinematical fields which are ``universal'' to $\I$:
In any asymptotically flat space-time, we obtain the same collection
of pairs. This situation is analogous to that in 4 dimensions where
pairs $(\t{q}_{ab}, \t{n}^a)$ constitute the universal kinematic
structure. However, whereas the 4-metric evaluated {\it at} $\I$ has
no dynamical content, in the present case, the 3-metric {\it at} $\I$
does carry dynamical content and varies from one space-time to another.

\goodbreak
\subsection{Asymptotic fields}
\label{s3.2}

The pairs $(\t{q}_{ab}, \t{n}^a)$ on $\I$ represent the leading or the
``zeroth oder'' structure at $\I$. The next, in the hierarchy, is an
intrinsic derivative operator. Let $\t{K}_b$ be a smooth co-vector
field on $\t{M}$, and $\u{\t{K}}_b$, its pull-back to $\I$. Define:
\be\t{D}_a \u{\t{K}}_b : =  \u{\t{\nabla}_a \t{K}_b} \  ,
\label{(3.9)} \ee
where the under-bar on right side denotes the pull-back to
$\I$. (Since $\u{\t{K}}_b = \u{\t{K}}'_b$ if and only if $\t{K}'_b =
\t{K}_b +\t{h} {}\t{n}_b + \Omega \t{W}_b$ for some smooth $\t{h}$ and
$\t{W}_b$, $\t{D}$ is a well-defined operator if and only if the
pull-back to $\I$ of $\t\nabla_a (\t{h} \t{n}_b + \Omega \t{W}_b)$
vanishes. It is easy to check that it does.) In 4 dimensions, the two
radiative degrees of freedom of the gravitational field are coded in
this intrinsic derivative operator \cite{18}. In 3 dimensions, on the
other hand, there is no ``pure'' gravitational radiation. Hence, one
would expect that the derivative operator $\t{D}$ has no invariant
physical content. This is indeed the case.

To see this, note first that given any vector field $\t{V}^a$
tangential to $\I$ we have:
$$ \t{V}^a\t{D}_a \t{q}_{ab} \= 0, \quad {\rm and}  \quad \t{V}^a
\t{D}_a \t{n}^b \= \t{V}^a\t{L}_a{}^b\, , $$
where, in the second equation, we have used Eq. (\ref{(3.5)}). Now,
for a zero rest mass scalar field (i.e., for 4-dimensional
Einstein-Rosen waves), $\t{L}_{ab} \= {\textstyle{1\over 2}}\t{\psi}^2
\t{n}_a \t{n}_b$, whence $\t{V}^a \t{D}_a \t{n}^a \= 0$. Hence, the
difference between any two permissible derivative operators on $\I$ is
given by:
$$ (\t{D}'_a - \t{D}_a) \t{K}_b \= \t{C}_{ab}^c \t{K}_c, \quad {\rm
with} \quad \t{C}_{ab}^c = \t\Sigma_{ab}\t{n}^c \t{K}_c\ ,\,\, $$
where $\t{K}_b$ is any co-vector field on $\I$ and $\t\Sigma_{ab}$, a
symmetric tensor field on $\I$, transverse to $\t{n}^a$;
$\t\Sigma_{ab}\t{n}^a \= 0$. Thus, $\t\Sigma_{ab} \= g\t{m}_a\t{m}_b$
for some function $g$ on $\I$. Now, if we make a conformal
transformation $\Omega \mapsto \Omega' = (1 + \omega_1
\Omega) \Omega$, the derivative operator $\t{D}$ changes through:
$(\t{D}'_a - \t{D}_a) \t{K}_b = \omega_1 \t{m}_a\t{m}_b \t{n}^c \t{K}_c$, {\it
even though the transformation leaves $\t{m}_a$ and $\t{n}^a$
invariant}. Thus, as in 4 dimensions, the ``trace-part'' of
$\t\Sigma_{ab}$ is ``pure-gauge''. Now, in 4 dimensions, the degrees
of freedom of the gravitational field reside in the trace-free part of
$\t\Sigma_{ab}$ \cite{18}. For the 3-dimensional description of
Einstein-Rosen waves, by contrast, since $\t\Sigma_{ab}$ is itself
pure-trace, the trace-free part vanishes identically reflecting the
absence of pure gravitational degrees of freedom.

In 4 dimensions, the Bondi news --which dictates the fluxes of
energy-momentum carried away by gravity waves-- is coded in the
curvature of $\t{D}$. By contrast, in the general 3-dimensional case
(i.e. without restriction on the form of matter sources), we can
always make the curvature vanish by going to an appropriate conformal
frame. To see this, recall, first that, since $\I$ is 2-dimensional,
the full curvature of any connection is determined by a scalar. For
connections under consideration, we have: $2\t{D}_{[a}\t{D}_{b]} \t{K}_c =
{}\t{R}\t\epsilon_{ab} \t{m}_c\t{n}^d \t{K}_d$, where $\t\epsilon_{ab}$ is the
obvious alternating tensor on $\I$. (Thus, $\t\epsilon_{ab} =
2\t{l}_{[a} \t{m}_{b]}$, where $\t{l}_a$ is a null co-vector field on $\I$
satisfying $\t{l}_a\t{n}^a = 1$.) Under conformal re-scalings $\Omega
{}\mapsto \Omega' = (1+\omega_1 \Omega)\Omega$, we have $\t{R} \mapsto
\t{R}' = \t{R}+ {\cal L}_{\t{n}}\omega_1$. Thus, by choosing an
appropriate $\omega_1$, we can always set $\t{R}' =0$. There is no
invariant physical information in the curvature of the derivative
operator $\t{D}$ intrinsic to $\I$.

Let us therefore examine the curvature of the full 3-dimensional
connection $\t\nabla$. Using Eq. (\ref{(3.1)}) and the Bianchi identity
of the rescaled metric $\t{g}_{ab}$ we have:
\be 2 \t{S}_{ab}\t{n}^a + \t\nabla_b(\Omega\t{f}) =
\t\nabla^a\t{L}_{ab} - \t\nabla_b \t{L} \, , \label{(3.10)}\ee
where $\t{L} = \t{g}^{ab}\t{L}_{ab}$.  The Bianchi identity for the
physical metric $g_{ab}$ implies that the right side of
Eq. (\ref{(3.10)}) is given by $ 2\Omega^{-1}
\t{L}_{ab}\t{n}^a$.  Hence, combining the two, we have:
\be 2 \t{S}_{ab}\t{n}^a + \Omega \t\nabla_b \t{f}
+ \t{f}\t{n}_b = 2 \Omega^{-1}\t{L}_{ab}\t{n}^a \, .\label{(3.11)}\ee
These, together with (3.1), are the basic equations that govern the
asymptotic dynamics.  

Our assumptions on the stress-energy tensor imply that $\Omega^{-1}
\t{L}_{ab}\t{n}^a\t{V}^b$ vanishes on $\I$ for any vector $\t{V}^a$ tangential
to $\I$.  Eq. (\ref{(3.11)}) now implies: $\t{S}_{ab} \t{n}^a \t{V}^b \=
0$.  Hence, the pull-back $\underbar{S}_{ab}$ to $\I$ of $\t{S}_{ab}$
has the form
$$ \underbar{S}_{ab} = \underbar{S} \t{m}_a\t{m}_b\ . $$
Similarly, since $\t{L}_{ab}$ is trace-free on $\I$ and since
$\t{L}_{ab}\t{n}^b$ vanishes there (cf. Eqs. (\ref{(3.3)}) and
(\ref{(3.4)})), the pull-back $\underbar{L}_{ab}$ of
$\Omega^{-1}L_{ab}$ to $\I$ exists and has the form:
$$ \underbar{L}_{ab} = \underbar{L} \t{m}_a\t{m}_b. $$
The field 
\be\t{B}:= \underbar{S} - \underbar{L} \label{(3.12)}\ee
will play an important role in what follows.

The Bach tensor $\t{B}_{abc}$ --vanishing of which is a necessary and
sufficient condition for conformal flatness in 3 dimensions-- is given
by:
\be \t{B}_{abc} = 2\t\nabla_{[b} \t{S}_{c]a} = 2\Omega^{-1}
(\t\nabla_{[b} \t{L}_{c]a} - \Omega^{-1}\t{n}^m\t{g}_{a[b}\t{L}_{c]m}).
\label{(3.13)}\ee
Thus, the Bach tensor is non-zero only in presence of matter. Note
that, in general, it does not vanish even at $\I$. This is in striking
contrast with the situation in 4 dimensions where the Weyl tensor of
the rescaled metric {\it does} vanish at $\I$. We will see that the
fact that in 3 dimensions we do not have conformal flatness even {\it
at} $\I$ makes the discussion of asymptotic symmetries much more
difficult. Transvecting the Bach tensor with $\t{n}^a$ and pulling the
result back to $\I$, we obtain:
\be\t{n}^a \u{\t{B}_{abc}} \= -{\cal L}_{\t{n}} 
\u{S}_{bc} \=  - {\cal L}_{\t{n}} \u{L}_{bc}
- (\lim_{\mapsto I} \Omega^{-2}\t{n}^m \t{n}^n \t{L}_{mn})
\t{q}_{bc}\, . \label{(3.14)}\ee
Since the last term in this equation has the form of the flux of
``matter-energy'' across $\I$ (it equals $2({\cal L}_{\t{n}}
\t\psi)^2$ in the case of Einstein-Rosen waves, cf. Eq. (\ref{se})), it
is tempting to interpret this equation as the analog of the local
Bondi conservation law on $\I$ in 4 dimensions. Let us rewrite this
equation in a more convenient form:
\be \t{D}_{[a}\,(\u{S} - \u{L}) \t{m}_{b]}
= {\textstyle{1\over 2}} \lim_{\mapsto \I}\, [\Omega^{-2} (\u{L}_{mn}
\t{n}^m \t{n}^n)\, \t\epsilon_{ab}]\, . \label{(3.15)}\ee
Then, it is tempting to regard the 1-form $\t{B}\t{m}_a \= (\u{S} -
\u{L}) \t{m}_{a}$ as the analog of the 4-dimensional ``Bondi 
mass aspect''. Let us therefore study its conformal properties. Under a
rescaling $\Omega \mapsto \Omega' = \omega\Omega$, we have:
\be \t{B}\t{m}_a \, \mapsto \, \t{B}'\t{m}'_a = 
[\omega^{-1} \t{B} - \omega^{-2}\t{m}^a\t{m}^b \t{D}_a\t{D}_b \omega  
+{\textstyle {3\over 2}}\omega^{-3} (\t{m}^a\t{D}_a
\omega)^2]\t{m}_a \, ,\label{(3.16)}\ee
where $\t{m}^a$ is a vector field tangential to $\I$ satisfying
$\t{m}^a \t{m}_a = 1$. Note that the transformation law involves only
the values of $\omega$ {\it on} $\I$; unlike in the transformation law
for $\t{R}$, discussed above, the field $\omega_1$ (which measures the
first derivative of $\omega$ off $\I$) never enters. This
transformation law will play an important role in the next two
sections.

Finally, we note an identity which enables us to express, at $\I$, the
quantity $\t{B}$ constructed from the curvatures of $\t{g}_{ab}$ and
${g}_{ab}$ in terms of the metric coefficients. To see this, recall
first that the derivative operator $\t{D}$ within $\I$ is obtained by
``pulling back'' the space-time derivative operator $\t\nabla$ to
$\I$. Hence one can express the curvature $\t{R}$ of $\t{D}$ in terms
of the curvature $\t{S}_{ab}$ of $\t\nabla$. Using the Bianchi
identity (3.10) to express some of the components of $\t{S}_{ab}$ in
terms of matter fields, we obtain:
\be\t{B} \, \= \, \underbar{S} - \underbar{L}\,  \= \,  
{\textstyle{1\over 2}} \t{f} - \t{R}\, . \label{(3.17)}\ee
Thus, in a conformal frame in which $\t{R}$ is zero, the analog
$\t{B}$ of the Bondi-mass aspect can be computed directly from the
metric coefficient $\t{f} = \Omega^{-2} \t{g}_{ab}\t{n}^a
\t{n}^b$. For the Einstein-Rosen waves, for example, it is
straightforward to check that the completion given in Sec.\ref{s2}
satisfies the condition $\t{R}= 0$ and by inspection $\t{f}$ is given
by $\exp\, (-2\t{\gamma})$. Thus, in practice, Eq. (\ref{(3.17)})
often provides an easy way to calculate $\t{B}$.  Finally, note that,
under conformal rescalings $\Omega \mapsto (1 + \omega_1\Omega)
\Omega$, both $\t{f}$ and $\t{R}$ transform non-trivially.  However, 
the combination $\textstyle \t{f} - \t{R}$ remains unchanged.

\goodbreak
\subsection{Point particle}
\label{s3.3}

In this sub-section, we will consider the simplest point-particle
solution to 3-dimension\-al gravity and, using the results obtained in
the last two sub-sections, study its behavior at null infinity.

In an obvious coordinate system adapted to the world line of the point
particle, the physical space-time metric $g_{ab}$ is given by \cite{19}:
$$ d\sigma^2 = -dt^2 + r^{-8GM}(dr^2 + r^2 d\phi^2), $$
where $-\infty < t< \infty,\, 0 <r < \infty $ and $0 \le \phi <
2\pi$. The particle has mass $M$ and ``resides'' at the origin. Since
the stress-energy tensor vanishes everywhere outside the $r=0$
world-line (which is excised from the space-time) the metric is flat
outside the origin. We can transform it in a manifestly flat form by
setting
$$ \rho = {r^\alpha\over \alpha}, \quad  \bar\phi = |\alpha|\phi,
\quad {\rm where}\quad \alpha = 1 -4GM \, .$$
(Note that $\bar\phi$ now ranges in $[0, 2\pi |\alpha|)$.) In terms of
these coordinates, the metric is given by:
\be d\sigma^2 = -dt^2 + d\rho^2 +\rho^2 d\bar\phi^2\, .  
\label{(3.18)}\ee
Although the metric is manifestly flat, it fails to be globally
Minkowskian because of the range of $\bar\phi$; there is a conical
singularity at the origin and the resulting deficit angle measures the
mass.

It is straightforward to conformally complete this space-time to
satisfy our definition of asymptotic flatness. Setting $u = t-\rho$
and $\Omega = 1/\rho$, the rescaled metric $\t{g}_{ab}$ is given by:
\be d\t{\sigma}^2 := \Omega^2 d\sigma^2 = - \Omega^2 du^2 
+ 2 du d\Omega + d\bar\phi^2\ . \label{(3.19)}\ee
It is trivial to check that the completion satisfies all our
conditions and that the conformal frame is divergence-free. The
kinematic fields are given by $\t{n}^a \equiv \partial/\partial u$ and
$\t{m}_a = \t{D}_a \bar\phi$. By inspection $\t{f} = 1$ and a
simple calculation shows that $\t{R} = 0$. Thus, $\t{B} = 1/2$; it
carries no information about mass. This information is hidden in the
deficit angle: Integrating $\t{m}_a$ on the base space ${\cal B}$, we
have:
$$ \oint_{\cal B} \t{m}_a dS^a\,  = \, 
2\pi \alpha = 2\pi (1-4GM)\, . $$

In 4 dimensions, one often insists that the conformal frame be such
that the metric on the base space be a unit 2-sphere metric. These are
the Bondi conformal frames. The obvious analog in 3 dimensions is to
ask that the frame be such that the length of the base space be equal
to $2\pi$, the length of a unit circle. (Although this restriction is
very weak, it seems to be the only viable analog of the Bondi
restriction in 4 dimensions.) The completion we gave above does not
satisfy this condition. However, it is trivial to rectify this
situation through a (constant) conformal rescaling. Set $\Omega' =
(1/\alpha)\Omega$. Then,
\be {d{\t{\sigma}}'}^2 = - {\Omega'}^2 du^2 + {2\over \alpha}du 
d\Omega'+ d \phi^2 \ , \label{(3.20)}\ee
where $\phi = (1/|\alpha|)\bar\phi$ ranges over $[0, 2\pi)$; the base
space ${\cal B}$ is a circle of length $2\pi$ as required. Since we
have performed a {\it constant} rescaling, we have $\t{R}' =
0$. However, $\t{f}$ does change: $\t{f}' = \alpha^2$. Thus, in the
``Bondi type'' frame, mass resides in $\t{B}$: Since $\t{B}=
{\textstyle{1\over2}}\alpha^2$ in this frame, the mass is given by 
\be M = {1\over 4G}(1 - \sqrt{2\t{B}})\, . \label{(3.21)}\ee
Thus, our expectation of the last sub-section that $\t{B}$ would be
the analog of the Bondi mass aspect is correct. However, to arrive at
this interpretation, we must use a properly normalized
(``Bondi-like'') conformal frame. This point will be important in
Sec.\ref{s5}.

We will conclude this discussion with two remarks. 

The metric considered in this sub-section is stationary and so it is
appropriate to compare the situation we encountered with that in
4-dimensional stationary space-times.  In both cases, the stationary
Killing field selects a preferred rest frame at $\I$ (which, in our
example, is given by the time translation $\partial /\partial{u}$).
However, in 4 dimensions, one can find {\it asymptotic} Killing fields
corresponding to space translations as well. In the present case, on
the other hand, due to the conical singularity, globally defined
space-translation vector fields fail to exist {\it even
asymptotically} (unless $M= 0$ in which case the deficit angle
vanishes). For example, we can introduce Cartesian coordinates $t,
\bar{x}, \bar{y}$ corresponding to $t,\rho,\bar\phi$. Then, $\bar{X}^a
\equiv \partial/\partial \bar{x}$ and $\bar{Y}^a \equiv
\partial/\partial \bar{y}$ {\it are} local Killing fields. However,
the chart itself fails to be globally defined and so do the vector
fields. Another strategy is suggested by what happens in Minkowski
space-time. In any of its standard completions space-translations are
represented by the vector fields $(\cos\phi) \t{n}^a$ and $(\sin\phi )
\t{n}^a$.  In the ``Bondi-like'' conformal frame introduced above
these vector fields are globally defined at null infinity of our point
particle space-time as well. However, now they fail to be Killing
fields even asymptotically. 

The second remark is that the stationary space-time we considered here
is a very special solution. Generic stationary solutions in
3-dimensional general relativity have a logarithmic behavior near
infinity and therefore fail to satisfy our definition of asymptotic
flatness at null infinity. (See Appendix B. Our point particle
solution corresponds essentially to the special case $C= 0$ in
Eqs. (\ref{B2},\ref{B3}).)  This is another key difference
between 3 and 4 dimensions. 

\goodbreak
\section{Asymptotic symmetries}
\label{s4}

In 4 dimensions, the asymptotic symmetry group at null infinity is
given by the BMS group \cite{13,15,17,30}. Its structure is the same as
that of the Poincar\'e group in that it is a semi-direct product of an
Abelian group with the Lorentz group. The Abelian group, however, is
{\it infinite} dimensional; it is the additive group of functions on a
2-sphere (the base space of $\I$) with conformal weight +1. It is
called the group of super-translations. The four dimensional group of
translations can be invariantly singled out. However, unless
additional conditions are imposed (near $i^0$ or $i^+$), the BMS group
does not admit a preferred Lorentz or Poincar\'e sub-group. This
enlargement from the ten dimensional Poincar\'e group to the infinite
dimensional BMS group is brought about because, in presence of
gravitational radiation, one can not single out a preferred Minkowski
metric even at infinity; one can only single out a family of
Minkowskian metrics and they are related by super-translations.

In this section, we will examine the asymptotic symmetry group in 3
dimensions. One's first impulse is to expect that the situation would
be completely analogous to that in 4 dimensions since the ``universal
structure'' available at $\I$ in the two cases is essentially the
same. It turns out however that because the space-time metric is
dynamical even at infinity --i.e., because in general the physical
metric does not approach a Minkowskian metric even to the leading
order-- the group of asymptotic symmetries is now enlarged even
further. Furthermore, now it is not possible to single out even the
group of translations without additional conditions.

This section is divided into two parts. The first discusses the
asymptotic symmetry group and the second introduces additional
conditions to single out translations.

\goodbreak
\subsection{Asymptotic symmetry group}
\label{s4.1}

Let us begin by recalling the universal structure, i.e., the structure
at infinity that is common to all asymptotically flat space-times. As
usual, the asymptotic symmetries will then be required to preserve
this structure.

Given {\it any} space-time satisfying our definition of asymptotic
flatness and {\it any} conformal completion thereof, its null
infinity, $\I$, is a 2-manifold, topologically $S^1\times R$. It is
ruled by a (divergence-free) null vector field $\t{n}^a$ and its
intrinsic, degenerate metric $\t{q}_{ab}$ satisfies:
\be\t{q}_{ab} \t{V}^b \= 0\ \ {\hbox{\rm if and only if}}\ \ 
\t{V}^b \propto \t{n}^b\, ,
\label{(4.1)}\ee
where $\t{V}^b$ is an arbitrary vector field on $\I$.  The ``base
space'' $\B$ of $\I$, i.e., the space of integral curves of $\t{n}^a$
on $\I$, has the topology of $S^1$. As in 4 dimensions, the intrinsic
metric $\t{q}_{ab}$ on $\I$ is the pull-back to $\I$ of a metric
$\bar{q}_{ab}$ on $\B$; that is, ${\cal L}_{\t{n}} \t{q}_{ab} = 0$.
Next, we have the conformal freedom given in Eq. (\ref{(3.2)}). Thus,
$\I$ is equipped with an equivalence class of pairs $(\t{q}_{ab},
\t{n}^a)$ satisfying Eqs. (\ref{(4.1)}, \ref{(3.5)}), where two are
considered as equivalent if they differ by a conformal rescaling:
$(\t{q}_{ab}, \t{n}^a)\approx (\omega^2 \t{q}_{ab}, \omega^{-1}
\t{n}^a)$, with $\L_{\t{n}} \omega = 0$. This structure is completely
analogous to that at null infinity of 4-dimensional asymptotically
flat space-times.

As we already saw, in 3 dimensions, a further simplification occurs:
in any conformal frame, $\I$ admits a unique co-vector field $\t{m}_a$
such that: $\t{q}_{ab} = \t{m}_a \t{m}_b$. Hence, in the universal
structure, we can replace $\t{q}_{ab}$ by $\t{m}_a$. Thus, $\I$ is
equipped with equivalence classes of pairs $(\t{m}_a,
\t{n}^a)$ satisfying:
\be \t{m}_a \t{n}^a \= 0 \quad {\rm and} \quad \L_{\t{n}} \t{m}_a \= 0 \ ,
\label{(4.2)}\ee
where $(\t{m}_a, \t{n}^a ) \approx (\omega \t{m}_a, \omega^{-1}
\t{n}^a)$ for any nowhere vanishing smooth function $\omega$ on $\I$
satisfying $\L_{\t{n}} \omega = 0$. Note that the second equation in
(4.2) implies that $\t{m}_a$ is the pull-back to $\I$ of a co-vector
field $\bar{m}_a$ on the base space $\B$.

The asymptotic symmetry group $\G$ is the sub-group of the
diffeomorphism group of $\I$ which preserves this structure. An
infinitesimal asymptotic symmetry is therefore a vector field
$\t{\xi}^a$ on $\I$ satisfying:
\be\L_{{\t\xi}} \t{m}_a \= \t\alpha \t{m}_a \quad{\rm and}\quad
\L_{\t\xi} \t{n}^a
\= - \t\alpha \t{n}^a\ , \label{(4.3)}\ee
for some smooth function $\t\alpha$ (which depends on ${\t\xi}^a$)
satisfying $\L_{\t{n}} \t\alpha \= 0$. Eqs. (\ref{(4.3)}) ensure that
the 1-parameter family of diffeomorphisms generated by ${\t\xi}^a$
preserves the ``ruling'' of $\I$ by the integral curves of its null
normal, its divergence-free character, and maps pair $(\t{m}_a,
\t{n}^a)$ to an equivalent one, thereby preserving each equivalence
class. It is easy to check that vector fields satisfying
Eqs. (\ref{(4.3)}) form a Lie algebra which we will denote by
$\L\G$. This is the Lie algebra of infinitesimal asymptotic
symmetries.

To unravel the structure of $\L\G$, we will proceed as in 4
dimensions. Let $\L\S$ denote the subspace of $\L\G$ spanned by vector
fields of the type ${\t\xi}^a \= \t{h} \t{n}^a$. Elements of $\L\S$
will be called infinitesimal {\it super-translations}. Eqs. (\ref{(4.3)})
imply:
\be\L_{\t{n}} \t{h} \= 0, \quad \L_{\t{h}\t{n}}\t{m}_a = 0,
\quad {\rm and}
\quad \L_{\t{h}\t{n}} \t{n}^a = 0\, .  \label{(4.4)}\ee
Thus, for any super-translation, $\t{h}$ is the pull-back to $\I$ of
$\bar{h}$ on the base space $\B$ and the action of the
super-translation leaves each pair $(\t{m}_a, \t{n}^a)$ individually
invariant. Furthermore, given any ${\t\xi}^a\in \L\G$ and any
$\t{h}\t{n}^a
\in \L\S$, we have:
\be [{\t\xi}, \t{h}\t{n} ]^a = (\L_{\t\xi} \t{h} - \t\alpha) 
\t{n}^a\ .\label{(4.5)}\ee 
Thus, $\L\S$ is a Lie ideal of $\L\G$.

To unravel the structure of $\L\G$, let us examine the quotient
$\L\G/\L\S$.  Let $[{\t\xi}^a]$ denote the element of the quotient
defined by ${\t\xi}^a$; $[{\t\xi}^a]$ is thus an equivalence class of
vector fields on $\I$ satisfying (4.3), where two are regarded as
equivalent if they differ by a super-translation.  The second equation
in (4.3) implies that every ${\t\xi}^a$ in $\L\G$ admits an
unambiguous projection $\bar{\xi}^a$ to the base space $\B$. The
equivalence relation implies that all vector fields ${\t\xi}^a$ in
$[{\t\xi}^a]$ project to the same field $\bar{\xi}^a$ on $\B$ and that
$[{\t\xi}^a]$ is completely characterized by $\bar{\xi}^a$. What
conditions does $\bar{\xi}^a$ have to satisfy? The only restriction
comes from the first equation in (4.3): $\bar{\xi}^a$ must satisfy
$\L_{\bar{\xi}}\bar{m}_a = \bar\alpha\bar{m}_a$ for some $\bar\alpha$
on $\B$. However, since $\B$ is {\it one} dimensional, this is no
restriction at all! Thus, $\bar{\xi}^a$ can be {\it any} smooth vector
field on the circle $\B$. $\L\G/\L\S$ is thus the Lie algebra of all
smooth diffeomorphisms on $S^1$. (In 4 dimensions, by contrast, the
first of equations (4.3) is very restrictive since the base space is a
2-sphere; $\bar{\xi}^a$ has to be a conformal Killing field on
$(S^2,\bar{q}_{ab})$. The Lie algebra of these conformal Killing
fields is just six dimensional and is isomorphic to the Lie algebra of
the Lorentz group in 4 dimensions.)

These results imply that the group $\G$ of asymptotic symmetries has
the structure of a semi-direct product. The normal subgroup $\S$ is
the Abelian group of super-translations. Given a conformal frame, each
infinitesimal super-translation ${\t\xi}^a = \t{h}\t{n}^a$ is characterized
by a function $\t{h}$. If we change the conformal frame, $\t{g}_{ab}
\mapsto \t{g}'_{ab} = \omega^2 \t{g}_{ab}$, we have $\t{n}^a\mapsto
\t{n}'^a = \omega^{-1}\t{n}^a$ and hence $\t{h}\mapsto \t{h}' = \omega
\t{h}$. Thus, each super-translation is characterized by a conformally
weighted function on the circle $\B$; the super-translation subgroup
$\S$ is isomorphic with the additive group of smooth functions on a
circle with unit conformal weight. The quotient $\G/\S$ of $\G$ by the
super-translation subgroup $\S$ is the group Diff$(S^1)$ of
diffeomorphisms on a circle.  In the semi-direct product, Diff$(S^1)$
acts in the obvious way on the additive group of conformally weighted
functions on $S^1$.

We will conclude this sub-section with some remarks. 

1. In the light of the above discussion, let us re-examine the
conditions on the stress-energy tensor in our definition of asymptotic
flatness. In Sec.\ref{s3.1} we pointed out that the conditions are
considerably weaker than those normally imposed in 4 dimensions and
argued that imposition of stronger conditions would deprive the
framework of interesting examples. Could we have imposed even weaker
conditions? Note that, if $\Omega T_{ab}$ fails to admit a
well-defined limit to $\I$, we could not even have concluded that $\I$
is a null hypersurface (see Eq. (\ref{(3.1)})). What about the
condition on the trace?  In absence of this condition, the pull-back
of $\t{L}_{ab}$ to $\I$ would not have vanished. This then would have
implied $\L_{\t{n}} \t{q}_{ab} \= (4/3)\t{L} \t{q}_{ab} \not=
0$. Consequently, the asymptotic symmetry group would have borne
little resemblance to the BMS group \cite{13,15,17,30} that arises in 4
dimensions. Thus, the specific conditions we used in the definition
strike a balance: they are weak enough to admit interesting examples
and yet strong enough to yield interesting structure at $\I$.

2. The semi-direct product structure of the asymptotic symmetry group
is the same as that of the BMS group. The super-translation group is
also the natural analog of the super-translation subgroup of the BMS
group. The quotient, however, is quite different: while it is the
Lorentz group in the 4-dimensional case, it is now an {\it infinite
dimensional} group, Diff$(S^1)$. Recall, however, that in the
corresponding analysis in 4 dimensions, the base space of $\I$ is a
2-sphere. $S^2$ admits a unique conformal structure and the Lorentz
group arises as its conformal group. In the present case, the base
space $\B$ is topologically $S^1$ and the quotient of $\G$ by the
super-translation subgroup is the conformal group of $S^1$. (Recall
that $\bar{\xi^a}$ has to satisfy $\L_{\bar{\xi}} \bar{q}_{ab} =
2\bar\alpha \bar{q}_{ab}$ since $\bar{q}_{ab} = \bar{m}_a\bar{m}_b$.)
It just happens that, since $S^1$ is 1-dimensional, {\it every}
diffeomorphism of $S^1$ maps $\bar{q}_{ab}$ to a conformally related
metric. This is the origin of the enlargement.

3. Can one understand this enlargement from a more intuitive 
standpoint? Recall that the symmetry group is enlarged when the
boundary conditions are weakened. Thus, it is the weaker conditions on
the fall-off of stress-energy --and hence on the curvature of the
physical metric-- that is responsible for the enlargement of the
group. This can be seen in the explicit asymptotic form of the metric 
of Einstein-Rosen waves that we encountered in Sec.\ref{s2.3},
\be d\sigma^2 =   e^{2\gamma} (-du^2 - 2dud\rho) + \rho^2 d\phi^2\  , 
\label{(4.7)}\ee 
where $\gamma \= \gamma(u)$ is a dynamical field on $\I$, sensitive to
the radiation. If $\gamma =0$, we obtain Minkowski space. The
radiative space-times that result when $\gamma \not= 0$ thus differ
from the radiation-free Minkowski space already to the {\it leading}
order at null infinity. In 4 dimensions, by contrast, the leading
order behavior of the physical metric has no dynamical content; the
components of the metric carrying physical information fall as $1/r$.
It is this difference that is responsible for the tremendous
enlargement of the asymptotic symmetry group.

Let us analyze this point further. Suppose, in  4 dimensions,
we consider metrics whose form is suggested by (4.6):
\be ds^2 = e^{2\gamma}(-du^2 -2 dudr) + r^2 d\Sigma^2 \ , 
\label{(4.8)}\ee
where $\gamma = \gamma(u,r,\theta, \phi)$ has a well defined limit as
$r$ tends to infinity along constant $u, \theta,\phi$ curves, and
$d\Sigma^2$ denotes the 2-sphere metric. Now, the situation is similar
to that encountered in the Einstein-Rosen waves: metrics with
different radiative content differ already to the leading
order. Nonetheless, setting $\Omega = 1/r$, it is easy to carry out a
conformal completion of this metric and verify that it admits a smooth
$\I$. However, the problem is that the {\it curvature of this metric
fails to fall off sufficiently rapidly for the stress-energy tensor to
have the fall-off normally required in 4 dimensions}. Hence, this
metric fails to be asymptotically flat in the usual 4-dimensional
sense. In 3 dimensions, on the other hand, to obtain an interesting
framework, we are forced to admit the analogous metrics (4.6).

\goodbreak
\subsection{Translations}
\label{s4.2}

In 4 dimensions, one can single out translations from the BMS group in
a number of ways. Somewhat surprisingly, it turns out that every one
of those techniques fails in 3 dimensions. We will first illustrate
this point and then show that one can introduce additional conditions
to single out translations. As one might expect from our discussion of
Sec.\ref{s3.3}, the situation is subtle even after introduction of
the stronger conditions.

Among various characterizations of the translation sub-group of the
BMS group, the one that is conceptually simplest and aesthetically
most pleasing is given by group theory \cite{30}: Translations form
the unique 4-dimensional {\it normal} subgroup of the BMS group. In
three dimensions, however, the asymptotic symmetry group is much
larger; the quotient of $\G$ by super-translations is now ${\rm
Diff}(S^1)$ --the {\it full} diffeomorphism group of a circle-- rather
than the (finite dimensional) Lorentz group. Consequently, $\G$ does
not admit {\it any} finite dimensional normal subgroup. Thus, the most
obvious 4-dimensional strategy is not applicable.

In 4 dimensions, another method of singling out translations is to use
the notion of ``conformal-Killing transport'' \cite{20}. The conformal
Killing data at any point of $\I$ corresponding to translations are
integrable because the Weyl tensor (of the tilde metric) vanishes
there identically.  In 3 dimensions, the analogous condition would be
vanishing of the Bach tensor. Unfortunately, as we saw in
Sec.\ref{s3.2}, in presence of matter fields the Bach tensor fails
to vanish at $\I$. (The explicit expression of the Bach tensor in the
case of Einstein-Rosen waves is given in Appendix A.) This in turn
makes the conformal-Killing transport of data that would have
corresponded to translations non-integrable on $\I$ and the strategy
fails.

Finally, a third method of selecting translations in 4 dimensions is
to go to a Bondi conformal frame, i.e., one in which the metric
$\bar{q}_{ab}$ on the base space is the unit 2-sphere metric and
consider the 4-parameter family of super-translations ${\t\xi}^a =
\t{h}\t{n}^a$, where $\t{h}$ is any linear combination of the $\ell= 0,
1$ spherical harmonics. There is only a 3-parameter family of Bondi
frames and the conformal factor that relates them is highly
constrained.  As a result, if $\t{h}$ is a linear combination of the
$\ell =0,1$ spherical harmonics in one Bondi frame, it is so in {\it
all} Bondi frames \cite{30}. The construction thus selects precisely a
4-parameter sub-group of the super-translation group $\S$.  This
strategy fails in 3 dimensions because the base space is now $S^1$ and
the notion of a ``unit $S^1$ metric'' fails to have the rigidity that
the unit 2-sphere metrics enjoy. Indeed, as we already remarked in
Sec.\ref{s3.3}, the only non-trivial analog of the Bondi frames
condition is to require that the conformal frame be such that the
length of the base space $\B$ be $2\pi$ and there is an {\it infinite}
dimensional freedom in the choice of such frames. Consequently, we can
not select a 3-dimensional space of translations in this manner.

Thus, to select translations, we need to impose additional conditions.
To be viable, they should select the standard, 3-dimensional
translation group in Minkowski space-time. However, as we saw in the
point particle space-time, asymptotic space-translations do not exist
globally near $\I$ if $M\not= 0$.  (This is also the case for
Einstein-Rosen waves.)  Hence, one would expect that, when the total
(ADM-type) mass is non-zero, the conditions should select only a time
translation. Thus, the conditions have to be subtle enough to achieve
both these goals at once. Fortunately, such conditions do exist and
are, furthermore, satisfied by a large class of examples.

A space-time $(M, g_{ab})$ will be said to be {\it strongly
asymptotically flat at null infinity} if it satisfies the boundary
conditions of Sec.\ref{s3.1} and admits a conformal completion in
which 
\be\t{B}\equiv {}\underbar{S} - \underbar{L} \equiv {1\over
2}\t{f} - {}\t{R} \,\, \rightarrow \textstyle{k\over 2} \ge 0
\quad \hbox{\rm as one approaches $i^o$ along $\I$\ ,} 
\label{saf}\ee
where $k$ is a constant.  Note that if the space-time is
axi-symmetric, $\t{B}$ automatically approaches a constant: if
$\Omega$ is chosen to be rotationally symmetric, $\t{B}$ would also be
rotationally symmetric everywhere on $\I$ and hence, in particular,
its limit to $i^o$ along $\I$ will be angle independent as
required. (We will see in Sec.\ref{s5} that the positivity of $k$
ensures that the ADM-type energy is well-defined.)  Thus, the
additional condition is satisfied in a large class of examples,
including the Einstein-Rosen waves and our point particle space-time.

Note that if the last condition is satisfied in a given conformal
frame, we can rescale the conformal factor by a {\it constant} and
obtain another conformal frame in which it is also satisfied. We can
eliminate this trivial freedom by a normalization condition. A
conformal frame will be said to be of {\it Bondi-type} if $\t{B}$
satisfies (\ref{saf}) {\it and} if $\oint_{\B} \t{m}_a dS^a = 2\pi$. A
natural question is: How many Bondi-type conformal frames does a
strongly asymptotically flat space-time admit? We will show that
Minkowski space admits precisely a 2-parameter family of them and the
freedom corresponds precisely to that of choosing a unit time-like
vector (i.e. a rest frame). This is completely analogous to the
freedom in the choice of Bondi frames in 4 dimensions. If the ADM-type
mass is non-zero, however, the Bondi-type frame will turn out to be
generically unique (unlike Bondi frames in 4 dimensions).

To establish these results, let us fix a strongly asymptotically flat
space-time and two Bondi-type completions thereof in which $\t{B}$ tends,
respectively, to $k/2$ and $k'/2$ for some constants $k$ and $k'$.
(In Minkowski space-time, it turns out that $k = k' =1$.) Let us
suppose that the two conformal frames are related by $\Omega
= \alpha \Omega'$, i.e., $\t{g}_{ab} = \alpha^2 \t{g}'_{ab}$. 
Then, the transformation property (3.17) of $\t{B}$ implies:
\be{k'\over 2} =  {k\over 2} \alpha^2 + \alpha\,
\t\partial^2 \alpha - {\textstyle{1\over 2}}\, 
(\t\partial\alpha)^2 \, ,
\label{(4.9)}\ee
where $\t\partial \equiv \t{m}^a\t{D}_a \equiv \partial/\partial\phi$.
The question now: How many (smooth) solutions does Eq. (\ref{(4.9)})
admit?  The equation is non-linear and rather complicated. However, if
we take its $\t\partial$-derivative we are left with a linear equation:
\be \t\partial (\t\partial^2 \alpha + k \alpha) = 0\, . 
\label{(4.10)}\ee 
This has regular solutions only if $k= n^2$ for an integer $n$ (recall
that, in a Bondi-frame, the range of $\phi$ on $\B$ is in $[0,
2\pi)$). Similarly, interchanging the role of primed and unprimed
frames, we conclude that $k' = {n'}^2$ for some integer $n'$. Finally,
the fact that the length of $\B$ in {\it both} conformal frames is
$2\pi$, implies that $n' = n$. Thus, unless $k=k' = n^2$,
Eq. (\ref{(4.9)}) does not admit a regular solution. Thus, unless $k=
n^2$, the Bondi-type conformal frame is in fact unique.  In this
generic case, we have a preferred time translation sub-group of $\G$
generated by ${\t\xi}^a = \t{n}^a$. In the point particle example,
this is precisely the time translation selected by the rest frame of
the particle. In Einstein-Rosen waves, it turns out to be the one
selected by the total Hamiltonian of the system \cite{10}.

If $k = n^2$, the reduced equation (\ref{(4.10)}) clearly admits a 
2-parameter family of solutions: In terms of the angular coordinate
$\phi$ on $\B$ (with $\t{m}_a = \t{D}_a\phi$), these are given by
\be \alpha = A + B \cos n\phi + C \sin n\phi, \quad {\rm with}
\quad - A^2 +B^2 +C^2 = -1\, . \label{(4.11)}\ee
It is straightforward to check that they also satisfy the full
equation (\ref{(4.9)}). 

In the obvious completion of Minkowski space-time (obtained by setting
$M = 0$ in the point particle example or $\psi =0$ in Einstein-Rosen
waves), we have $\t{f}=1$ and $\t{R} = 0$, whence $\t{B} = 1/2$. This
corresponds to the case $n =1$. Thus, Minkowski space-time does admit
Bondi-type conformal frames and the constant $k$ is precisely $1$
(i.e., we can not obtain any other value by going from one Bondi-type
frame to another). There is precisely a 2-parameter family of
Bondi-type frames related by a conformal factor $\alpha$ of
Eq. (\ref{(4.11)}) (with $n=1$). Fix any one of these and consider the
3-parameter family of super-translations of the form $\t{h}\t{n}^a$
where
\be \t{h} = (a + b \cos\phi + c \sin\phi)\, . \label{(4.12)}\ee
Using Eq. (\ref{(4.11)}) (with $n = 1)$, one can check that this
3-dimensional space of these super-translations is left invariant if
we replace one Bondi-type frame by another. Following the (third)
strategy (mentioned above) used in 4 dimensions, one can call this the
translation sub-group of the asymptotic symmetry group. This label is
indeed appropriate: It is easy to check that the restrictions to $\I$
of any translational Killing field of Minkowski space has precisely
this form. Thus, if $n=1$, the procedure does select for us a
3-dimensional translation sub-group of $\G$.

It turns out, however, that if $n =1$, the deficit angle at spatial
infinity vanishes and we therefore have zero ADM-type energy. By
3-dimensional positive energy theorem \cite{10}, the only physically
interesting space-time in which this can occur is the Minkowski
space-time.  If $k >1$, we have a surplus angle at spatial infinity
and the ADM-type energy is now negative.  We will therefore ignore the
$n>1$ cases from now on (although they do display interesting
mathematical structures; see Appendix B). 

To summarize, strongly asymptotically flat space-times generically
admit a preferred Bondi-type frame and a preferred
time-translation. In the exceptional cases, where $k = n^2$, we obtain
a 3-parameter family of Bondi-type frames. However, the only
physically interesting exceptional case is Minkowski space-time where
$n=1$.

\goodbreak
\section{Conserved quantities}
\label{s5}

This section is divided into two parts. In the first, we introduce the
notion of energy at a retarded instant of time and of fluxes of energy
and, in the second, we discuss super-momenta. Again, while the general
ideas are similar to those introduced by Bondi, Sachs and Penrose in 4
dimensions, there are also some important differences.

Perhaps the most striking difference is the following. Consider
generic, strongly asymptotically flat space-times. As we saw, in this
case, there is a preferred Bondi-type frame and a preferred
translation subgroup of the asymptotic symmetry group. However, as the
example of Einstein-Rosen waves illustrates, because the space-time
metric is dynamical even at infinity, the vector field $\t{n}^a$ (or a
constant multiple thereof) in the Bondi-type frame is {\it not} the
extension to $\I$ of a {\it unit} time translation in the space-time.
If the initial data of the scalar field are of compact support,
space-time is flat in a neighborhood of $i^o$ and a constant multiple
of $\t{n}^a$ --namely, $(\exp \, \t\gamma_0) \t{n}^a$-- coincides with
the extension to $\I$ of the unit time translation near
$i^o$. However, in the region of $\I$ with non-trivial radiation, the
restriction of the unit time translation is given by $(\exp\,\t\gamma
(u)) \t{n}^a$; the rescaling involved is $u$-dependent whence the
vector field is not even a super-translation!  Energy, on the other
hand, is associated with unit time translations. Hence, energy at null
infinity is not directly associated with any component of
super-momentum and a new strategy is needed to define it.

\goodbreak
\subsection{Energy}
\label{s5.1}

The strategy we will adopt is to capture the notion of energy through
the appropriate deficit angle. We will first begin with motivation,
then write down the general expression of energy and finally verify
that it has the expected physical properties.

Let us begin with an axi-symmetric, strongly asymptotically flat
space-time, consider its Bondi-type completion with an axi-symmetric
conformal factor. (Thus, $\oint_{\B} \t{m}_a dS^a = 2\pi$.) Fix a
cross-section $C_o$ of $\I$ to which the rotational Killing field is
tangential. Because of axi-symmetry of the construction, the field
$\t{B}$ is constant on $C_o$, say $\t{B}|_{C_o} = k_o/2$. If this were
a cross-section of $\I$ of the point particle space-time, it follows
from our discussion of Sec.\ref{s3.3} (cf Eq. (\ref{(3.21)})) that we
would associate with it energy
\be E= {1\over 4G}(1 -\sqrt{k_o})\, .\label{(5.1)}\ee
(Thus, in particular, if $k_o= 1$ as in Minkowski space-time, we would
have $E= 0$.) 

By inspection, we can generalize this expression to arbitrary
cross-sections of null infinity of general --i.e., non-axi-symmetric--
space-times. Given any strongly asymptotically flat space-time, a
Bondi-type conformal frame and a cross-section $C$ of $\I$, we will
set:
\be E[C] := {1\over 8\pi G}\oint_{C}\, (1 - \sqrt{2\t{B}})\,\t{m}_a 
dS^a \, .\label{(5.2)}\ee
The appearance of the square-root is rather unusual and seems at first
alarming: the formula would not be meaningful if $\t{B}$ were to
become negative. Note, however, that, by assumption of strong
asymptotic flatness, the limit $k/2$ of $\t{B}$ to $i^o$ is
positive. Furthermore, since ${\cal L}_{\t{n}} \t{B} =
\lim_{\mapsto\I} \,\Omega^{-2}\, \t{L}_{cd}\t{n}^c \t{n}^d$ and
since the right side is positive definite if the matter sources
satisfy local-energy conditions, $\t{B}$ remains positive on
$\I$. Thus, $E[C]$ is bounded above by $1/4G$ which is also the upper
bound of the total Hamiltonian at spatial infinity \cite{10}.
 
Let us now verify various properties of this quantity which provide a
strong support in favor of its interpretation as energy.

{}$\bullet$ First, let us suppose that we are in Minkowski
space-time. Then, in {\it any} Bondi-type frame, we have $\t{B} = 1/2$
everywhere on $\I$. Hence, on any cross-section, the energy vanishes.

{}$\bullet$ Next, let us consider the point-mass space-time with
positive $M$. Then from Sec.\ref{s4.2} we know that there is a unique
Bondi-type frame and in this frame, ${2\t{B}} = (1- 4GM)^2$ whence,
on {\it any} cross-section $C$, we obtain $E[C] = M$. This is of
course not surprising since our general definition was motivated by
the point mass example. However, the result is not totally trivial
because we are now allowing arbitrary cross-sections, not necessarily
tangential to the rotational Killing field.

{}$\bullet$ Consider Einstein-Rosen waves. In the non-trivial case when
the scalar field $\psi$ is non-zero, the Bondi-type frame is unique.
In this frame, $2\t{B} = \exp (-2\t\gamma (u))$. Hence,
$$ E[C] = {1\over 8\pi G}  \oint_C (1 - e^{-\t\gamma (u)})d\phi\, .$$
In the limit to $i^o$ (or, in the past of the support of $\t\psi (u)$
on $\I$), we have $E \mapsto (1/4G)(1- \exp {(-\t\gamma_0}))$. This is
{\it precisely} the value of the total Hamiltonian at spatial infinity
--the generator of unit time translations near $i^o$. This result is
non-trivial because the Hamiltonian is defined \cite{10} through {\it
entirely} different techniques using the symplectic framework based on
Cauchy slices.  In the limit to $i^+$, we know from Sec.\ref{s2.3}
that $\t\gamma (u)$ tends to zero. Hence $E[C]$ tends to zero. This
behavior of $E[C]$ is also physically correct because $i^+$ is regular
in these space-times. We wish to emphasize that these two constraints
--agreement with the known expressions both at $i^o$ and $i^+$ of
Einstein-Rosen waves-- on the viable expression of energy are
strong. Hence, the fact that there exists a {\it general} expression
for $E[C]$ involving only fields defined {\it locally} on the cross-section
 $C$ which reduces to the correct limits at both ends of $\I$
of the Einstein-Rosen waves is quite non-trivial.

{}$\bullet$ What about the flux of energy? If a cross-section $C_1$ is
in the future of a cross-section $C_2$, from Eqs. (\ref{(3.15)},
\ref{(5.2)}) we have:
\ba E[C_1] - E[C_2] &=& {1\over 8\pi G} \int_{\Delta} 
{\t{D}}_{[a} (1 - \sqrt{2\t{B}})\, \t{m}_{b]} dS^{ab}\nonumber\\
&=& -{1\over 16\pi G} \int_{\Delta}\, (2\t{B})^{-{1\over 2}}\,
\lim_{\mapsto I} (\Omega^{-2} \t{L}_{mn}\t{n}^m \t{n}^n)\, 
\t\epsilon_{ab} dS^{ab}\, ,\label{(5.3)}\ea
where $\Delta$ is the portion of $\I$ bounded by $C_1$ and $C_2$.
(The limit in the integrand is well-defined because of our conditions
on the stress-energy tensor. For the Einstein-Rosen waves, it is
$(\L_{\t{n}}\t{\psi})^2$; see Eq. (\ref{se}).)  If the matter sources
satisfy local energy conditions, the integrand in the second
expression is positive definite. Thus, $E[C_1] \le E[C_2]$, the
equality holding if and only if there is no flux of matter through the
region $\Delta$. As one would expect, radiation through $\I$ carries
positive energy. The appearance of $1/\sqrt{2\t{B}}$ in the integrand
is not alarming because, as remarked above, for the class of
space-times under consideration, $\t{B}$ is guaranteed to be positive
on $\I$ in Bondi-type frames.

{}$\bullet$ In the case when the source is a zero rest-mass scalar
field, we can make the energy flux more explicit: $\lim_{\mapsto I}
(\Omega^{-2} \t{L}_{mn}\t{n}^m \t{n}^n) = 2 ({\cal L}_{\t{n}}
\t\psi)^2$.  Hence, for Einstein-Rosen waves, Eq. (5.3) reduces to:
\be E[C_1] - E[C_2] = -{1\over 8\pi G} \int_{\Delta}\, 
e^{\t\gamma(u)} ({\cal L}_{\t{n}}\t{\psi})^2\, \t{\epsilon}_{ab} 
dS^{ab}\, .  \label{(5.4)}\ee
In the limit in which the cut $[C_2]$ tends to $i^o$, $E[C_2]$ reduces
to the gravitational Hamiltonian \cite{10}. Hence, on any cut, $E[C]$
is given by the difference between the total Hamiltonian and the
energy that is radiated out up until that cut. Finally, note that,
because of the appearance of $\exp \t\gamma(u)$ in the integrand, this
expression of energy-flux is more complicated than the flux formula
(\ref{(2.37)}) for $\gamma(u)$, i.e., the flux formula for Thorne's
C-energy \cite{2}. This is, however, to be expected: Even at spatial
infinity, the total Hamiltonian is $(1/4G)(1-\exp (-\t\gamma_o))$
while the C-energy is just $(1/4G)\t\gamma_o$. In the weak field limit
the two agree. But in strong fields, they are quite different. In
particular, the total Hamiltonian and $E[C]$ are bounded above by
$1/4G$ while the C-energy is unbounded above.

{}$\bullet$ We saw that, in the case of Einstein-Rosen waves, our
expression (5.2) of energy reduces to the total Hamiltonian in the
limit as the cross-section approaches $i^o$. We expect that this
result holds much more generally: It should hold in any space-time
which is strongly asymptotically flat at null infinity and also
satisfies the boundary conditions at spatial infinity needed in the
Hamiltonian formulation \cite{10}. That is, broadly speaking, we
expect the agreement to hold if the space-time is sufficiently
well-behaved to have a well-defined total Hamiltonian {\it and} a
well-defined limit of (5.2) to $i^o$. It is easy to provide strong
plausibility arguments for this conjecture since both quantities
measure the deficit angle at $i^o$. However, more detailed arguments
are needed to establish this result conclusively.

\goodbreak
\subsection{Super-momentum}
\label{s5.2}

We will conclude the main paper by introducing a notion of
super-momentum. For reasons indicated in the beginning of this
section, however, these quantities are not related to the energy in a
simple way. They are given primarily for completeness.  As in 4
dimensions \cite{18}, in a suitable Hamiltonian formulation based on
null infinity, they may be the generators of canonical transformations
induced by super-translations.

Recall first that, in 4 dimensions, super-momentum arises as a linear
map from the space of super-translations to reals and is expressible
in any conformal frame. The basic fields that enter are constructed
from the asymptotic curvature of the rescaled metric (and matter
sources).  However, in order to ``remove irrelevant conformal factor
terms'', one also has to introduce a kinematic field \cite{17} with
appropriate conformal properties. The situation in 3 dimensions is
rather similar.

Let us begin by introducing the analog $\t\rho$ of the kinematical
field.  Set $\t\rho = 1/2$ in any Bondi-type conformal frame and
transform it to any other frame via the following law: if $\Omega
=\alpha \Omega'$, then
\be{\t{\rho}}' = \alpha^2 \t\rho + \alpha \, \t\partial^2\alpha - 
{\textstyle{1\over 2}} (\t\partial \alpha)^2\, , \label{(5.5)}\ee
where, as before $\t\partial \equiv \t{m}^a \t{D}_a$. Hence, the field
$\t\rho -\t{B}$ transforms rather simply: $({\t\rho}' - {\t{B}}') =
\alpha^2 (\t\rho - \t{B})$ (see Eq. (3.17)). As in 4 dimensions,
the field $\rho$ serves two purposes: it removes the unwanted,
inhomogeneous terms in the transformation properties of $\t{B}$ and 
it removes the ``purely kinematical'' part of $\t{B}$ in the Bondi-type
frames.

We can now define the super-momentum. Fix any conformal completion
of the physical space-time (not necessarily of a Bondi-type). The 
value of the super-momentum on a super-translation $\t{T}\t{n}^a$,
evaluated at a cross-section $C$ of $\I$ will be:
\be P_{\t{T}}[C] = {1\over 8\pi G} \oint_{C}\,(\t\rho - \t{B})\, 
\t{T} \t{m}_a dS^a\ . \label{(5.6)}\ee 
Under a conformal transformation, $\Omega \mapsto \Omega'=\alpha^{-1}
\Omega$, we have ${\t{T}}' = \alpha^{-1} \t{T}$ and ${\t{m}_a}' = 
\alpha^{-1}\t{m}_a$.  Hence, the 1-form integrand remains unchanged.
Thus, as needed, the expression of super-momentum is conformally
invariant; i.e., it is well-defined. 

Let us note its basic properties. First, by inspection, the map
defined by the super-momentum $P$ from super-translations to reals is
linear. Second, in Minkowski space-time, $\t\rho = \t{B}$ in any
conformal frame. Hence, the value of super-momentum vanishes
identically on {\it any} cross-section. Finally, since ${\cal
L}_{\t{n}} \t\rho = 0$, we have
\be{\cal L}_{\t{n}} [(\t\rho - \t\B)\t{T} \t{m}_a] = 
- \lim_{\mapsto I} (\Omega^{-2}\,\t{L}_{mn}\t{n}^m
\t{n}^n)\t{T} \t{m}_a\, .\label{(5.7)}\ee
Therefore, as in the case of energy, the flux of the component of the
super-momentum along any time-like super-translation (i.e., one in
which $\t{T} > 0$) is positive.

\goodbreak
\section{Discussion}
\label{s6}

In this paper, we developed the general framework to analyze the
asymptotic structure of space-time at null infinity in 3 space-time
dimensions. We did not have to restrict ourselves to any specific type
of matter fields. However, if the matter sources are chosen to be a
triplet of scalar fields constituting a non-linear ($SO(2,1)$)
$\sigma$-model, the space-times under considerations can be thought of
as arising from symmetry reduction of 4-dimensional generalized
cylindrical waves, i.e., vacuum solutions to the 4-dimensional
Einstein equations with one space-translation isometry. If
the source consists of a single zero rest mass scalar field, the
translation Killing field in four dimensions is hypersurface
orthogonal. Finally, if there is, in addition, a rotational Killing
field, the space-times are symmetry reductions of the 4-dimensional
Einstein-Rosen waves.

The general strategy we adopted was to follow the procedures developed
by Bondi and Penrose in 4 dimensions. However, we found that
due to several peculiarities associated with three dimensions, those
procedures have to be modified significantly. A number of unexpected
difficulties arise and the final framework has several surprising
features. This is in contrast with the situation in higher dimensions
where the framework is likely to be very similar to that in 4
dimensions.

The new features can be summarized as follows. First, in 3 dimensions,
the space-time metric is flat in any open region where stress-energy
vanishes and thus we are forced to consider gravity coupled with
matter. To accommodate physically interesting cases, we have to allow
matter fields such that the fall-off of the stress-energy tensor at
null infinity is significantly weaker than that in 4 dimensions. This
in turn means that the metric is dynamical even at infinity; it does
not approach a Minkowskian metric even in the leading order. In fact,
physically interesting information, such as the energy and
energy fluxes, is coded in these leading order, dynamical terms. As a
result, the asymptotic symmetry group $\G$ is enlarged quite
significantly. Like the BMS group in 4 dimensions, it admits an
infinite dimensional normal subgroup $\S$ of super-translations. The
structure of this sub-group is completely analogous to that of its
counterpart in 4 dimensions. However, the quotient, $\G/\S$, is {\it
significantly} larger. While in 4 dimensions the quotient is the six
dimensional Lorentz group, now it is the infinite dimensional group
${\rm Diff}(S^1)$ of diffeomorphisms of a circle. Furthermore, whereas
the BMS group admits a preferred (4-dimensional) group of
translations, $\G$ does not. To select translations, one has to impose
additional conditions, which in some ways are analogous to the
conditions needed in 4 dimensions to extract a preferred Poincar\'e
subgroup of the BMS group. We imposed these by demanding that there
should exist a conformal frame in which the field $\t{B}$ tends to a
constant as one approaches $i^o$ along $\I$. This condition is
automatically satisfied in axi-symmetric space-times. We saw that, in
a generic situation, it selects a unique conformal frame (up to
constant rescalings which can be removed by a normalization condition)
and we can then select a preferred time translation in $\S$.  If the
past limit of the $\I$-energy is zero, it selects a 2-parameter family
of frames ---the analogs of Bondi frames in 4 dimensions. In this
case, we can select a 3-dimensional sub-group of translations from
$\S$. Finally, given any cross-section $C$ of $\I$, we associated with
it energy, $E[C]$, as well as a super-momentum $P_{\t{T}}[C]$. The
former is a scalar and has several properties that one would expect
energy to have. The latter is a linear map from the space of
super-translations to reals and may arise, in an appropriate
Hamiltonian formulation based on $\I$, as the generator of canonical
transformations corresponding to super-translations.

These results refer to 3-dimensional general relativity coupled to
arbitrary matter fields. However, as noted above, if the matter fields
are chosen appropriately, we can regard the 3-dimensional system as
arising from a symmetry reduction of 4-dimensional vacuum general
relativity by a space-translation Killing field. (One can also
consider 4-dimensional general relativity coupled to suitable
matter. Then, one acquires additional matter fields in 3 dimensions.)
In this case, the energy $E[C]$ (or the super-momentum $P_{\t{T}}[C]$)
associated with a cross-section $C$ of 3-dimensional null infinity
represents the energy (or super-momentum) per unit length (along the
symmetry axis) in four dimensions. Thus, the 3-dimensional results
have direct applications to 4-dimensional general relativity as well.
In addition, as we will see in the companion paper \cite{16}, the
analysis of the asymptotic behavior of fields in 3 dimensions can also
be used to shed light on the structure of  null infinity in 4 dimensions.

There are a number of technical issues that remain open. First, as
indicated in Sec.\ref{s5.1} it is desirable to find the precise
conditions under which the past limit of $E[C]$ yields the total
Hamiltonian \cite{10}. A second important issue is that of positivity
of $E[C]$. For the total Hamiltonian, this was established \cite{10}
using a suitable modification of Witten's spinorial argument in 4
dimensions.  Can this argument be further modified to show positivity
of $E[C]$?  If space-time admits a regular $i^+$, the limit of $E[C]$
as $C$ tends to $i^+$ vanishes. Since the flux is positive, this
implies that $E[C]$ is positive on every cross-section. However, in
the general case, it is not apriori clear that in the Bondi-type
frame, $\t{B}$ will not exceed $1/2$ making $E[C]$ negative on some
cross-section. Next, in the case when the matter fields admit initial
data of compact support, space-time is flat near $i^o$. In this case,
it should be possible to select a preferred 1-parameter sub-group of
rotations in $\G$ and define angular momentum. Finally, in the case
when $i^+$ is regular, one would expect that, as in Minkowski space,
there exists a 2-parameter family of Bondi-type conformal frames in
which $\t{B}$ tends to a constant at $i^+$. It is not apriori clear
whether the Bondi-type frame selected by the behavior of $\t{B}$ at
$i^o$ is included in the family selected at $i^+$. If the space-time
is axi-symmetric, the answer is in the affirmative. It would be
interesting to investigate what happens in the general case.

The present framework provides a natural point of departure for
constructing an $S$-matrix theory both classically and, especially,
quantum mechanically. 3-dimensional quantum gravity without matter
fields is fully solvable but the solution is trivial in the
asymptotically flat case. When we bring in matter, we have a genuine
field theory which is diffeomorphism invariant. If the matter fields
are suitably restricted, the theories are equivalent to the reduction
of 4-dimensional general relativity (or of 10-dimensional string
theories).  Quantization of such theories should shed considerable
light on the conceptual problems of non-perturbative quantum
gravity. As a first step towards quantization, one might use ideas
from the asymptotic quantization scheme introduced in 4 dimensions
\cite{21}. Since the Lorentz sub-groups are now replaced by the ${\rm
Diff}(S^1)$ sub-groups of ${\cal G}$ and since ${\rm Diff}(S^1)$
admits interesting representations (with non-zero central charges),
the asymptotic quantum states would now have interesting, non-trivial
sectors. Secondly, this quantization would also lead to ``fuzzing'' of
space-time points along the lines of Ref. \cite{22}. To see this,
recall first that the light cone of each space-time point gives rise
to a ``cut'' of $\I$ (which, in general, is quite complicated). Thus,
given $\I$ and these light cone cuts, one can ``recover'' space-time
points in an operational way. Now, in a number of cases with scalar
field sources --including of course the Einstein-Rosen waves-- one
expects the initial-value problem based on $\I$ to be well-posed and
the classical $S$-matrix to be well-behaved.  In such cases, it should
be possible to express the light cone cuts on $\I$ directly in terms
of the data of the scalar field on $\I$. Now, in the quantum theory,
the scalar field on $\I$ is promoted to an operator-valued
distribution and, given any quantum state, one only has a probability
distribution for the scalar field to assume various values. This
immediately implies that one would also have only probability
distributions for light cone cuts, i.e., for points of
space-time. This approach may well lead one to a non-commutative
picture of space-time geometry.
\bigskip\goodbreak

{\bf Acknowledgements:} AA and JB would like to thank the
Max-Planck-Institute for Gravitational Physics for its kind
hospitality. AA was supported in part by the NSF grants 93-96246 and
95-14240 and by the Eberly Research Fund of Penn State University. JB
was supported in part by the grants GACR--202/96/0206 and
GAUK--230/1996 of the Czech Republic and the Charles University, and
by the US-Czech Science and Technology grant 92067.
\bigskip\goodbreak

\appendix
\section{Riemann and Bach tensors}

In this Appendix we will provide the behavior of the Riemann and Bach 
tensors at null infinity in (2+1) dimensions.

Assume the metric to be given in Bondi--type coordinates $ (x^0, x^1,
x^2) = (u, \rho , \phi)$ as in (\ref{(2.26)}). The Christoffel symbols
are $$
\Gamma^0_{00}= 2\gamma_{,u}-\gamma_{,\rho}\ , \ \ \ \ \Gamma^0_{22}=
\rho\  e^{-2\gamma}\ ,
$$
$$
\Gamma^1_{00} = \gamma_{,\rho}-\gamma_{,u}\ , \ \ \  \Gamma^1_{01}=
\gamma_{,\rho}\ , \ \ \Gamma^1_{11} =2\gamma_{,\rho}\ .
$$
\be
\Gamma ^1_{22}=-\rho\ e^{-2\gamma}\ , \ \ \ \ \ \Gamma^2_{12}= 
\rho^{-1}\ .\ee
The Riemann tensor ($R^i_{jkl}=\Gamma^i_{jl,k}-\dots$) reads
$$
R_{0101}=e^{2\gamma}(\gamma_{,\rho\rho}-2\gamma_{,u\rho})\ ,\ \ 
R_{0202}=\rho\ (\gamma_{,\rho}-\gamma_{,u})\ ,\
$$
\be R_{0212}=\rho\ \gamma_{,\rho},\ \ \ R_{1212}=2\rho\ 
\gamma_{,\rho}\ .\ee

In a general (2+1)--dimensional spacetime the Riemann tensor has
the form 
\be  R_{ijkl}=2(S_{i[k}g_{jl]}-S_{j[k} g_{il]})\ , \ee
where 
\be S_{ik}= R_{ik}-{1\over 4}g_{ik}R\ . \label{A4}\ee
It has six independent components given by the symmetric tensor
$S_{ik}$. In the case of the rotation symmetry the following
components are non-vanishing: 
$$ 
S_{00}= {1\over 2} \gamma_{,\rho\rho}-\gamma_{,u\rho}
+\rho^{-1}(\gamma_{,\rho}-\gamma_{,u})\ , $$ $$ S_{01}=S_{10}={1\over
2} \gamma_{,\rho\rho}-\gamma_{,u\rho} +\rho^{-1}\ \gamma_{,\rho}\ , 
$$
$$ 
S_{11}=2\rho^{-1}\ \gamma_{,\rho}\ , 
$$ 
\be  S_{22}= \rho^2\ e^{-2\gamma}({1\over 2} \gamma_{,\rho\rho}-
\gamma_{,u\rho})\ . \ee

The role of the Weyl tensor in 3 dimensions is played by the
conformally invariant Bach tensor (see e.g. \cite{1}): 
\be B_{ijk}=S_{ik;j}-S_{ij;k}\ . \ee
The Bach tensor satisfies $B_{ijk}=-B_{ikj}$ and $B_{[ijk]cykl}=0$,
and it thus has five independent components. In the
rotation--symmetric case the Bach tensor writes ($\delta
=\gamma_{,\rho}-2\gamma_{,u}$) 
$$ 
B_{001}= {1\over
2}(\delta_{,u}-\delta_{,\rho})_{,\rho}+(\delta +\gamma_{,u})
(\delta_{,\rho}+\rho^{-2}) -\rho^{-1}\delta_{,\rho}\ , $$
$$B_{101}=-{1\over 2}\delta_{,\rho\rho}+(\delta +2\gamma_{,u})
(\delta_{,\rho}+\rho^{-2}) -\rho^{-1}\delta_{,\rho}\ , $$ 
\be B_{202}= \rho^2e^{-2\gamma}(B_{001}-B_{101}) \ ,\ \ \ 
B_{212}= -\rho^2e^{-2\gamma}B_{101}\ .
\ee
Let us choose the real null triad
\be l^i=(0,e^{-2\gamma},0)\ ,\ n^i=(1,-{1\over 2},0)\ , \ \ 
m^i=(0,0,\rho^{-1})\ .
\ee
It is easy to see that it is parallel propagated along $u=$const,
$\phi=$const, and it satisfies 
\be l_in^i=-1\ ,\ m_im^i=1\ , \
l_il^i=l_im^i=n_in^i=n_im^i=0\ .
\ee
Further, let us introduce six real triad components of the Riemann
tensor or, equivalently, of the tensor $S_{ik}$ given by (A4) as
follows: 
$$ 
{\it S}_1 = R_{ijkl}l^im^jl^km^l=S_{ik}l^il^k\ , 
$$ 
$$
{\it S}_2 = R_{ijkl}l^in^jl^km^l=S_{ik}l^im^k\ , 
$$ 
$$ {\it S}_3 =
R_{ijkl}({1\over 2}l^in^jl^kn^l- m^in^jm^kl^l)=S_{ik}m^im^k\ , 
$$ 
$$
{\it }S_4 ={1\over 2} R_{ijkl}l^in^jl^kn^l=S_{ik}l^in^k\ , 
$$ 
$$ 
{\it S}_5 = R_{ijkl}n^il^jn^km^l=S_{ik}n^im^k\ , 
$$ 
\be {\it S}_6 = R_{ijkl}m^in^jm^kn^l=S_{ik}n^in^k\ .
\ee
Under the rotation symmetry we find
$$
{\it S}_1=2\rho^{-1}\gamma_{,\rho}e^{-4\gamma}\ , \ \ {\it S}_2=0\ ,
$$
$$
{\it S}_3={\it S}_4=e^{-2\gamma}({1\over 2} \gamma_{,\rho\rho}-
\gamma_{,u\rho})\ , \ \ {\it S}_5=0\,\ , 
$$
\be {\it S}_6=\rho^{-1}(\gamma_{,\rho}-\gamma_{,u})\ .
\label{A11}\ee
Assume now the scalar field admits an expansion (\ref{(2.23)}). The
field equations (\ref{(2.27)}), (\ref{(2.28)}) imply 
$$\gamma_{,u}= -2\dot f_0^2 -{1\over 2}  f_0\dot f_0 {1\over\rho} 
+\dots\ ,
$$
\be \gamma_{,\rho}={1\over 4} f_0^2 {1\over\rho^{2}}+ \dots\ .
\ee
The Riemann tensor (\ref{A11}) has then the following
asymptotic form: 
$$ 
{\it S}_1={1\over 2}e^{-4\gamma_\infty}f_0^2
{1\over {\rho^3}}+ O({1\over {\rho^4}})\ , 
$$ 
$$ {\it S}_3={\it
S}_4={1\over 2}e^{-2\gamma_\infty}f_0\dot f_0 {1\over {\rho^2}}+
O({1\over {\rho^3}})\ , 
$$ 
\be  {\it S}_6=2\dot
f_0^2{1\over\rho}+O({1\over{\rho^2}})\ ,
\ee
where $\gamma_\infty=\lim_{\rho\to\infty} \gamma (u,\rho)$.

Finally, define the five real triad components (scalars) of 
the Bach tensor:
$$
{\it B}_1=B_{ijk}l^in^jm^k\ , \ \ {\it B}_2=B_{ijk}l^il^jm^k\ ,\ \ 
{\it B}_3=B_{ijk}n^in^jm^k\ ,
$$
\be {\it B}_4=B_{ijk}m^im^jl^k\ , \ \ \ {\it B}_5=B_{ijk}m^im^jn^k\ .
\ee
Under the rotation symmetry we find only the last two scalars 
non-vanishing. Their asymptotic behavior is the following:
$$
{\it B}_4=-{1\over 4}e^{-4\gamma_\infty}\left[6(f_0f_1)\dot{} + 
f_0^3\dot f_0 \right]{1\over{\rho^4}} + O({1\over{\rho^5}})\ ,
$$
\be {\it B}_5={1\over 2}e^{-2\gamma_\infty}\left[f_0\ddot f_0 
-3\dot f_0^2 +4f_0\dot f_0^3
\right]{1\over{\rho^2}} + O({1\over{\rho^3}})\ .
\ee

Now the Bach tensor is conformally invariant and it is of interest to
see precisely its form at null infinity in the unphysical spacetime.
Putting $\tilde\rho=\rho^{-1}$, $\tilde u=u$, $\tilde\phi=\phi$
and using again $\Omega=\tilde\rho$ as in (\ref{(2.33)}), we
introduce the null triad in the unphysical space by $\tilde
l=\Omega^{-2}l, \tilde n=n, \tilde m=\Omega^{-1}m$, so that in the
coordinates ($\tilde u,\tilde\rho, \tilde\phi$) we have 
\be \tilde l^i=(0,-e^{-2\tilde\gamma}, 0)\ , \ 
\tilde n^i=(1,{1\over 2}\tilde\rho^2,0)\ , \ \tilde
m^i=(0,0,1)\ .\ee
(Note that the vector $\t{n}^i$ is null everywhere. Outside $\I$, it
is not related in any simple way to the vector field $\t{n}^a :=
\t{g}^{ab}\t\nabla \Omega$ used in the main text.)

Using then $\tilde B_{ijk}=B_{ijk}$ we arrive at the following form of
the Bach tensor at null infinity {\it I} :
$$
{\it\tilde B}_4=\tilde B_{ijk}\tilde m^i\tilde m^j\tilde l^k=
-{1\over 4}e^{-4\tilde\gamma_0}\left[
6(\tilde f_0\tilde f_1)_{,\tilde u}+\tilde f_0^3 \tilde f_{0,\tilde u} 
\right]+O(\tilde\rho)\ , 
$$
\be {\it\tilde B}_5=\tilde B_{ijk}\tilde m^i\tilde m^j\tilde n^k=
{1\over 2}e^{-2\tilde\gamma_0}\left[
\tilde f_0 \tilde f_{0,\tilde u\tilde u}-3 \tilde f _{0,\tilde u}^2
+4\tilde f_0 \tilde f_{0,\tilde u}^3 \right]
+O(\tilde\rho)\ ,\ee
where $\tilde\gamma_0=\tilde\gamma(\tilde u,\tilde\rho=0)=
\gamma_\infty$, $\tilde f_0(\tilde u)=f_0(u)$, $\tilde
f_1(\tilde u)=f_1(u)$. Hence the Bach tensor is finite and
non-vanishing at null infinity in general.
\goodbreak

\section{Asymptotics for static cylinders in 3 dimensions}

Starting from the 4-dimensional Einstein-Rosen metric,
\be d{} s^2 = e^{2{}\gamma - 2{}\psi} (-dt^2+d{} 
\rho^2) + e^{2{}\psi} d{} z^2 + {}\rho^2 e^{-2{} 
\psi} d{} \phi^2 \; , \label{B1}\ee
Marder \cite{23} gives 4-dimensional static solution representing the
field outside a static cylinder in the form: 
\ba \psi &=& -C(1-C)^{-1} \ln {} \rho - (1-C) \ln D \; ,
\label{B2}\\ 
\gamma &=& C^2(1-C)^{-2} \ln {} \rho - (1-2C) \ln D \; , 
\label{B3}\ea
where $C$ and $D$ are constants which can be determined, by matching
the solution to an interior one, in terms of mass and pressure
distribution inside the cylinder. For mass $M$ per unit length of the
cylinder small, Levi-Civita and others suggest that $C=2M$; Thorne's
C-energy \cite{2} leads to the same results as long as the internal
pressure of the cylinder is much smaller than its energy density.

The simplest models of the static cylinders employ thin shells. By
studying the exterior and flat interior metric of an infinite static
cylindrical shell, Stachel \cite{24} found the constants $C$ and $D$ to be
related to the internal structure of the cylinder in a simple
way. Denoting the radius of the shell by $\rho_0$, and introducing
Stachel's notation, $a$ and $A^+$, for the constants determining the
external metric, we find Marder's constants $C$ and $D$ to be given by
$$
C={a\over {a-1}} \ ,
$$
\be \ln D ={1-a\over {1+a}}(a^2\ln\rho_0 + \ln A^+)\ ,\ee
so that 
\be \gamma=a^2\ln {\rho\over \rho_0} - \ln A^+ \ ,\ee
\be \psi=a\ln {\rho\over \rho_0} + b\ .\ee
An additive constant $b$ in $\psi$ can be removed by a rescaling $\rho
\to \xi\tilde\rho\ ,t \to \xi\tilde t\ , z \to \xi^{-1}\tilde z\ , 
\psi \to \tilde\psi + \ln\xi\ ,  \gamma\to \tilde\gamma$,
$\xi = const$, which leaves the metric (\ref{B1}) invariant.

Let $S_{ab}$ be the surface stress--energy tensor of the shell. Then
Stachel's equations (1.7 a,b,c) determine the surface energy density,
$\sigma=S_t{}^t$, and the surface pressures, $p_z=-S_z{}^z$,
$p_\phi=-S_\phi{}^\phi$, in terms of the constants $a$ and $A^+$ as
follows:
$$
\sigma = {1-A^+\over\rho_0}\ ,
$$
\be p_z={A^+(a-1)^2-1\over \rho_0}\ \ \ ,\ \ p_\phi = {a^2A^+\over
\rho_0}\ . \ee
The dominant energy condition, $\sigma\ge 0$, $|p_z|,\ |
p_\phi|\le\sigma$, requires
\be 1-A^+\ge0\ \ \ ,\ \ -\left [{1-A^+\over A^+}\right ]^{1\over 2} 
<a\le 0\ . \ee
Choosing $a=0$, $0<A^+<1$, we obtain the cylinders with 
\be \sigma={1-A^+\over \rho_0} =- p_z\ \ ,\ \ p_\phi=0\ ,\ee
generating the exterior fields as straight cosmic strings: locally
flat but conical, with a positive deficit angle given by $
2\pi(1-A^+)$. Curiously, if we admit a negative mass density such that
\be A^+=1+n\ ,\ \ n=1\ ,2\ , \dots\ , \label{B10}\ee
and thus
\be \sigma= -{n\over\rho_0}\ \ = -p_z\ , \ee
the exterior space is some covering space of a part of Minkowski
space. Indeed, it is easy to see that with $\gamma=-\ln (1+n)$, $\psi
= const$, the metric (\ref{B1}) can be converted to a flat
metric with $\bar\phi\in [0,2\pi(n+1)]$. The holonomy group of such a
space is the same as that of a part of Minkowski space so that vectors
transported parallel around closed curves coincide with the original
( cf. also \cite{26} and \cite{27} who find no ``gravitational
Aharonov--Bohm effect" in the cases corresponding to $A^+$ given by
(\ref{B10})). The Lie algebra of Killing fields does not differ
from that of a part of Minkowski space. However, the geometry
(determined by the metric itself, rather than by the connection) is
different. With the original coordinate $\phi\in[0,2\pi)$ it reads
(after rescaling $t$) 
$$ 
ds^2 =-dt^2+ {1\over (n+1)^2} d\rho^2 + \rho^2 d\phi^2 + dz^2.  
$$ 
Considering surfaces $t = const$, $ z = const$, and comparing the
proper lengths, $2\pi\rho_1$ and $2\pi\rho_2$, of the two circles with
radii $\rho_1$ and $\rho_2$, with their proper ``orthogonal distance",
$(n+1)^{-1}(\rho_2 -\rho_1)$, the result differs from that in
Minkowski space. This (anti)conical character of spacetime can be
observed also at infinity after performing an inversion using
Cartesian coordinates (cf. Eqs. (2.19) of \cite{16}). This, of course,
is true for any (anti)conical space with $A^+\ne1$.

In any case, the asymptotic gravitational field describing static
cylinders is determined by two parameters, rather than one, describing
the asymptotic field of cylindrical waves considered in the main text.
(Relatively recently, Bondi \cite{25} examined quasi-statically changing
cylindrical systems and concluded that there is no conservation of
these parameters because of gravitational induction transferring
energy parallel to the axis.)

The (2+1)-dimensional metric corresponding to (\ref{B1}) is
(cf. (\ref{(2.11)}))
\be d\sigma^2 = e^{2{} \gamma} (-d{} t^2 + d{} 
\rho^2) + \rho^{2} d{} \phi^2 \; . \ee
Introducing ${} u = {} t - {} \rho$ and writing ${} 
\gamma$ in the form
\be {} \gamma = a^2\ln {} \rho + B \; , \quad a^2 \ge 0 \; ,\ 
B~\hbox{constants} \; , \ee
we get
\be d{}\sigma^2 = {} \rho^{2a^2} e^{2B} (-d{} u^2 - 2d{} 
u d{} \rho) + {} \rho^2 d{} \phi^2 \; . \ee
Now we go over to the unphysical 3-dimensional spacetime with 
coordinates
\be \tilde u = {} u \; , \quad \tilde\rho = {} \rho^{2a^2-1} \; , 
\quad 
\tilde\phi = {} \phi \label{B15}\ee
by a conformal transformation with the conformal factor
\be \Omega = \tilde\rho^{-1/(2a^2-1)}\ . \label{B16}\ee
The metric of the unphysical spacetime then reads 
\be d\tilde\sigma^2 = \Omega^2 d\sigma^2 = e^{2B}
\left[-\tilde\rho^{2(a^2-1)/(2a^2-1)} d\tilde u^2 - 2(2a^2-1)^{-1}
d\tilde u d\tilde\rho\right] + d\tilde\phi^2 \; . \label{B17}\ee
Assume $a^2<{1 \over 2}$. This includes cases when mass per unit
length of the cylinder is small because then constant $C\ll 1$ and
$0<a^2=C^2 (1- C)^{-2} \ll 1$. Transformation (\ref{B15}) shows that
${} \rho \to \infty$ implies $\tilde\rho \to 0$, and (\ref{B16})
implies $\Omega = 0$ at $\tilde\rho = 0$. The metric (\ref{B17})
becomes degenerate here. The conformal completion of the spacetime
with a given $a^2<{1 \over 2}$ can thus be constructed, with infinity
being at $\Omega = 0$. However, (\ref{B16}) yields $\nabla\Omega = 0$
at $\tilde\rho = 0$. Therefore, the asymptotics for static cylinders
is completely different from a standard conformal completion of an
asymptotically flat spacetime. In special cases of locally flat but
conical space-times the asymptotics in (3+1)--dimensional context is
analyzed in \cite{28}.
\goodbreak

\end{document}